\documentclass[a4paper,12pt]{article}

\usepackage{bbm}
\usepackage[utf8]{inputenc}
\usepackage{placeins}
\usepackage{amssymb,amsmath,amsfonts}
\usepackage[colorlinks,linkcolor=purple,citecolor=teal]{hyperref}
\usepackage{dsfont}
\usepackage{color}
\usepackage{graphicx}
\usepackage{hyphenat}
\usepackage{wrapfig}
\usepackage{empheq}
\usepackage{textcomp}
\usepackage[caption=false]{subfig}
\usepackage{rotating}
\usepackage{wrapfig}
\usepackage{pdfpages}
\usepackage{verbatim}
\usepackage{setspace}
\usepackage{array}
\usepackage{caption}
\usepackage[pdf]{pstricks}
\usepackage{upgreek}
\usepackage{cmll}
\usepackage{latexsym}
\usepackage{braket}
\usepackage{cite}
\usepackage{epsfig}
\usepackage[left=2.4cm,top=3.3cm,right=2.4cm,bottom=3.3cm,bindingoffset=0cm]{geometry}
\usepackage[titletoc,page]{appendix}
\usepackage{multicol}
\usepackage{youngtab}

\newcommand{\beq}{\begin{eqnarray}}
\newcommand{\eeq}{\end{eqnarray}}
\newcommand{\bea}{\begin{eqnarray}}
\newcommand{\eea}{\end{eqnarray}}
\newcommand{\be}{\begin{equation}}
\newcommand{\ee}{\end{equation}}

\def\brc{\langle}
\def\ckt{\rangle}

\def\Tr{\qopname\relax o{Tr}}

\numberwithin{equation}{section}

\numberwithin{equation}{section}

\begin{document}
 

\title{
Quantum fluctuations, particles and  entanglement: \\
a discussion towards the solution of   \\
the quantum measurement problems \footnote{Dedicated to the memory of Giampiero Paffuti.   }  }


\author{  
 Kenichi Konishi$^{(1,2)}$   \\[13pt]
 {\em \footnotesize
$^{(1)}$INFN, Sezione di Pisa,    
Largo Pontecorvo, 3, Ed. C, 56127 Pisa, Italy}\\[2pt]
{\em \footnotesize
$^{(2)}$Department of Physics ``E. Fermi", University of Pisa,}\\[-5pt]
{\em \footnotesize
Largo Pontecorvo, 3, Ed. C, 56127 Pisa, Italy}\\[2pt]
\\[1pt] 
{ \footnotesize  kenichi.konishi@unipi.it  }  
}
\date{}


\maketitle

\begin{abstract}

 The quantum measurement problems are revisited from a new perspective. 

  One of the main ideas of this work is that the basic entities of our world are various types of particles, elementary or composite. 
 It follows that  each elementary process, hence each measurement process at its core, is a spacetime,  pointlike, event. 
  Another key idea is that, when a  microsystem  $\psi$  gets into contact with the experimental device,  factorization of $\psi$  rapidly fails and entangled mixed states appear.
 The wave functions  for the microsystem-apparatus coupled systems for different measurement outcomes then lack overlapping spacetime support. 
It means that  the aftermath of each measurement  is a single term in the sum:  a ``wave-function collapse". 

Our  discussion leading  to a diagonal density matrix, $\rho= {\rm diag} ( |c_1|^2, \ldots, |c_n|^2, \ldots )$ shows how 
the information encoded  in the wave function    $|\psi\ckt    =    \sum_n  c_n   | n \ckt$  gets transcribed,  via entanglement with the experimental device and environment, into the relative frequencies   ${\cal P}_n = |c_n|^2$  for various experimental results $F=f_n$.   

These results  represent new, significant steps  towards filling in the logical gaps in the  standard 
interpretation based on Born's rule,  and replacing it with a more natural one.  Accepting objective reality of quantum fluctuations, independent of any experiments, and independently of human presence,  one renounces
  the idea that in a fundamental, complete theory of Nature 
the result of each  single experiment must necessarily be predictable. 

A few well-known puzzles such as the  Schr\"odinger cat  conundrum and the EPR paradox
are  briefly reviewed:
 they can all be naturally explained away.

\end{abstract}

\bigskip

\newpage

\tableofcontents

\newpage

\section{Introduction   \label{Intro} }  

The  so-called  ``quantum measurement problem" has been with us ever since the establishment of quantum mechanics in the first quarter of the last century. 
The validity of quantum mechanics has since passed every tests made, first through an impressive success in  atomic physics.  
Today, we may note that 
the standard $SU(3)_{QCD} \times (SU(2)\times U(1))_{GW\!S}$ model  of the fundamental  (strong and electroweak) interactions,   based on the relativistic quantum mechanics of the particles and fields, 
is one of the most successful and precisely tested  physics theories known so far.  
We should mention also many beautiful quantum mechanical phenomena in condensed matter physics, such as superfluidity, superconductivity, quantum Hall effect, Bose-Einstein condensation of ultracold atoms, and so on.

In spite of all this, the probabilistic nature of the quantum mechanical predictions has always kept 
 us with an uneasy feeling, that something fundamental is missing in our understanding of quantum mechanics. 
 
 Such a feeling is often expressed in the form of  a ``quantum measurement problem".  It actually represents several different issues.  
  Is the wave-function-collapse which apparently occurs
at the moment of a measurement   real?  Is the macroscopic superposition of the states involving the microsystem, the experimental apparatus, and eventually the environment (and in principle the rest of the 
world), relevant?   Are we living in a section of the forever branching many-worlds tree?  Or does the whole world spontaneously collapse (or localize)  from time to time?   Or is the wave function just an (extremely clever) bookkeeping device which is nothing truly physical but that somehow encodes the information relating past experimental results to future ones? Together with  these philosophical ``problems", there  is  a conceptual  conflict between the smooth and deterministic  time evolution of the systems  described by the Schr\"odinger equation,  and the quantum jumps occurring regularly  in atoms, nuclei and molecules, or at the moments of the measurements.  And above all,  is  Born's rule  something which follows in principle  from the Schr\"odinger equation involving  everything from the microsystem, the measuring apparatus, and the whole world, via some mysterious  effectively nonlinear evolution? And independently of all this, is it conceivable that a complete, fundamental theory of Nature 
should not be able to predict the result of each single experiment uniquely?
   \footnote{A concise  but systematic  discussion of these  ``quantum measurement problems"  can be found
 in Chap. 19 and Sec.~23.1 of  \cite{KKGP}. Many earlier references (up to 1983) are in  Wheeler and Zurek \cite{WheelerZ}. 
 Many related issues are discussed also in Peres' book \cite{Peres}. 
See  Bell \cite{Bell} for some in-depth discussions and further references. }
 
  This work addresses all of these questions and puzzles.   The careful reader will find a clear answer or resolution to each of them,
 in the following discussions. 
 Our answer to the last question is 
throughout the work: it is summarized in the Abstract and in Sec.~\ref{Conclusion}.

The actual  measurement devices can vastly differ in their nature, size, materials used, and technologies employed,  from a simple photographic plate, cloud and bubble  chambers filled with liquid or vapour, spark chambers and MWPC made 
with metal plates, wires and gas,  the neutrino detectors made of a huge tank of pure water and tens of thousands of  photomultipliers,  to the state-of-the-art  silicon detectors and some future apparatus which uses superconducting materials for dark matter search. 
 Such an enormous diversity of the experimental devices requires  that  an equally vast simplification be made, to capture the essence of   quantum measurements.   

We know  that, independent of the details, 
a good experimental device  faithfully reflects the quantum fluctuations of the system
 being studied, described by the wave function $\psi$.  Indeed, the result of each single experiment is, in general, apparently random and unpredictable  \footnote{An exception occurs when the state $\psi$
 is  one of the eigenstates of the quantity $F$, with $F= f_m$,  i.e.,   $|\psi\ckt = |m \ckt$.   In such a case, a good experiment  produces the same result $F=f_m$  every time.
  }.    
   Yet,  the information contained in the wave function $\psi$,
   encoded as the expectation values  for any variable  \footnote{Throughout, we use the word ``variable",  for  a dynamical variable, physical quantity, or  observable,    such as energy, momentum, angular momentum, position, etc. }    $F$ (and all functions thereof)  
    \be       {\bar F}  \equiv     \brc  \psi | F  | \psi \ckt \;,     \label{fund}  
 \ee
 fully manifests itself, as a (frequency-)  average  of the experimental outcomes,    $  {\bar  {f_n }}$,      
 \be    {\bar {f_n}} =   \sum_n       f_n  \,  {\cal P}_n   \;, \qquad    {\cal P}_n  =     |\brc  n  | \psi \ckt|^2 \;,  \label{notBorn}  
 \ee
 where  $F |n\ckt = f_n | n \ckt$, and  ${\cal P}_n$ is  the relative frequency  for  $f_n$.  
 Such a  prediction of quantum mechanics is verified 
  by countless experiments.

The prediction (\ref{notBorn})  would follow,  if one assumes  that  the  {\it  probability } of  each single measurement of $F$ 
  in  the state $\psi$ to give $f_n$,     is given by  
$    P_n  =     |\brc  n  | \psi \ckt|^2 \;,$
i.e.,  Born's rule.   This assumption is presented in most textbooks as one of the fundamental postulates  of quantum mechanics 
\footnote{A rare exception is the famous book by  Dirac  (1958)  \cite{Dirac},  
     where slightly difference nuance is used.   }.

 We introduce here a slight change of perspective. The  fundamental laws of quantum mechanics  are  represented by the expectation values (quantum fluctuation average)  of   all 
 variables in the system.   Such quantum fluctuations are there, independently of any experiments.       As will be illustrated and discussed in Sec.~\ref{General},  Sec.~\ref{measurement} and Sec.~\ref{Conclusion},
       the new  point of view constitutes a  small but nonetheless significant conceptual departure from the traditional  way of interpreting  the fundamental laws of quantum mechanics. 

The main results of this work (Sec.\,\ref{Main})  follow from the following two  key ideas. 
The first  ([A]  below)  which  leads to the resolution of most of the puzzles and paradoxes associated with the measurement processes,   comes from the observation that  the basic building blocks of our world  are various types of {\it particles}.
 The elementary constituents of Nature (the electron, the muon, the photon, the quarks, the $W, Z$ bosons, the gluons, etc.) are all pointlike objects (particles), and all the fundamental interactions  occur at definite  spacetime points,   the vertices of the Feynman diagrams.

    It follows that  [A]   
   {\it   at its heart   each  experiment 
   is a   space-time pointlike event, or, triggered by one}  \footnote{The specification ``or triggered by"  is important, as  some measurement, such as  a momentum measurement, involves a 
   sequence of such events - the particle tracks. 
   See Sec.~\ref{WFcollapse}.  }.
Such an observation was at the core of  Einstein's explanation of the photoelectric effects based on Planck's light quantum.  

One might object that  
 certain static or adiabatic  effects such as the Coulomb potential describing the atomic structures, the 
  magnetic fields used in the Stern-Gerlach experiment or in a momentum measurement,  and  the electric and magnetic fields used to accelerate particles
  in high-energy experiments,  
  do not represent  themselves spacetime pointlike events. These are particular effects of quantum electrodynamics, the quantum theory  of particles (i.e.,  electrons and  photons).     
Nevertheless,  these effects do not inject energy to the  microscopic system in interaction with them, suddenly enough 
 to trigger  the chain ionization processes, hadronic cascades, and the amplification and the consequent recording    of the particle fingerprints  by the device,  
 characteristic of a measurement process.
This fact  allows us to use these static  or adiabatic  effects in many useful ways,
   while  at the same time keeping them  from becoming {\it  a
measurement} themselves.  This discussion hopefully clarifies, partially at least,  the question  which often arises:   which part of the processes should be regarded as  ``the measurement" and which part not   (e.g. a preparation) 
  \footnote{To be complete, and having in mind a possible further objection from a particle theorist,  let us note that even a static effect  (e.g. a constant electric field),  if strong enough,  can create particle pairs from the vacuum  by quantum effects  (known as  the vacuum polarization \cite{Schwinger}). These processes  are eventlike,   but in general are not directly related to the measurement process under consideration here. }.

We also note that the molecules, atoms, nuclei, and even particles such as proton, neutron and pions  (which were once considered as  elementary particles) are
all bound states (i.e., composite particles) made of some other, underlying constituents. This fact does not affect the statement of the preceding paragraphs:  all these composite  objects  behave perfectly as pointlike particles at appropriate 
distance scales, much  larger than their characteristic sizes.  
 
Taking into account this eventlike nature of the measurement processes \footnote{The emphasis given here on the space-time, eventlike nature of each measurement, might  remind some reader of the ``ETH-approach to Quantum Mechanics" \cite{Frohrich}; 
however,  the concept of  ``events"  used there is different from ours. We use the word  ``event" in the ordinary sense of the word.     }
 will indeed be central in showing the absence of interferences among  the distinct  states of  the microscopic quantum and  macroscopic apparatus  combined system, which is one of the principal issues in the ``quantum measurement problems".  See the discussions in Sec.~\ref{measurement} $\sim$
 Sec.~\ref{entanglement} below.

 A second  main input of this work  ([B] below) concerns the central role played by factorization  and entanglement  (see Sec.~\ref{factorization})  in the measurement processes.
 The concept of pure state, e.g., of the wave function $|\psi\ckt$  describing the microsystem before the measurement,  
 is a (usually very) good approximation,  if the experiment is carefully prepared.  Still, it describes a subsystem of a larger system containing the experimental apparatus and the rest of the world.  Exact factorization  means that the wave function of the whole system   before the measurement   has a factorized form   
 \be      |\Psi\ckt =    |\psi\ckt \otimes   |\Phi; Env\ckt\; \label{factorizationpsi}
 \ee
 (see Sec.~\ref{measurement} for the definition of   the apparatus-environment coupled state, $ |\Phi; Env\ckt$).
The second key idea  of this work  [B] is  that,     
  during  the measurement process,   
factorization of $|\psi\ckt$, as in  (\ref{factorizationpsi}),  rapidly fails as a result of interactions  between it and the measurement device and with the rest of the world,   and  as a result 
  an entangled  mixed state labelled by the unique measurement reading  emerges,  see  (\ref{step111}), (\ref{diagonal})  below.    
   The result is  a state-vector reduction, often called  a  ``wave-function collapse"
   \footnote{In the author's view,  ``wave-function collapse" is one of the worst misnomers  in the quantum mechanics discussions.  The words evoke in our mind a mysterious, nonlinear evolution which shrinks almost instantaneously whatever distribution present in $\psi$  before the measurement.   No such processes exist.  Every experimentalist knows what happens in each measurement:    they are  chain-ionizations, hadronic cascades in a calorimeter, recording of the particle tracks,  etc.   However complicated in detail,   they are all processes we understand well in principle,   in terms of  the standard theory of the fundamental interactions.  What happens is an effective spacetime localization of each measurement event, as explained in
   Sec.~\ref{loss}  below.}.

 The main result of this work (summarized in Sec.\,\ref{Main}) which follows from 
the analyses of Sec.~\ref{General} $\sim$ Sec.~\ref{measurement}, is  
 the well-known quantum mechanical prediction, 
  that measurements done on the state $\psi$
  \be    | \psi \ckt  =     \sum_n   c_n  \, |n \ckt\;, \qquad  F | n\ckt = f_n |n \ckt\;, 
\ee
   give various results  $F=f_n$, with   relative frequencies ${\cal P}_n = |c_n|^2$.
This may sound almost identical to  the standard  postulate of quantum mechanics:  
  ``a  measurement done on the state $\psi$ will give various results  $F=f_n$, with probabilities, $P_n = |c_n|^2$"  (Born's rule).

What is missing in  Born's rule, however,  is the {\it   explanation of how }  the information about  the quantum fluctuations encoded in the wave function  $\psi$ 
 gets transcribed, through the measurement process,  into the expected relative frequencies  of finding various results $F= f_n$,  ${\cal P}_n = |c_n|^2$, and {\it the understanding of  why and how}   the state-vector reduction takes place, after each experiment. 
 
 The contribution of this work is to indicate the first steps to  fill in  these  gaps.

 In Sec.~\ref{jump} $\sim$ Sec.~\ref{entanglement}  we discuss  several  well-known  puzzles and apparent  paradoxes, and show how they  are all naturally explained.
 Sec.~\ref{Universe} discusses a few general ideas about the universe  and the quantum measurements,  which come out of our considerations. 
We  summarize and conclude   in Sec.~\ref{Conclusion}.

\section{The expectation values versus  Born's rule \label{General} }

We first review the equivalence of  the information contained in the expectation values (the quantum fluctuation averages)  and Born's rule.

\subsection {The angular momentum and spin  \label{Spin} }

In an  angular momentum-spin system, the quantum fluctuations are famously    expressed by the fact that 
in a state of definite angular momentum magnitude  ${\mathbbm J}^2 = j(j+1)$, each component of   ${\mathbbm J}$ can attain  only up to the maximum absolute  value 
$j$,   which is smaller  (for $j \ne 0$) than   the square root of  ${\mathbbm J}^2$.  This is due to  the fact that  the three components $J_x, J_y, J_z$ do not commute with each other: they are not mutually compatible \footnote{We will
use, throughout,  the same symbol (such as $F$ and $J_x$)    to indicate a physical variable itself,  the mathematical  self-adjoint operator representing it, and sometimes even the (eigen-) values 
these variables take.    The alert reader however should  not have any difficulty  telling which is meant, each time.}.
In a state   with definite  $({\mathbbm J}^2,  J_z)$,   the values of  $J_x$ and $J_y$ are undetermined -  Heisenberg's uncertainties.
Quantitatively,  in the state   (we set $\hbar =1$)
\be      {\mathbbm J}^2  | j, m \ckt = j(j+1) | j, m \ckt\;; \quad    J_z | j, m \ckt = m | j, m \ckt\;,    \qquad    m=j, j-1, \ldots, -j\;,  
\ee
 the values of   $J_x$ and $J_y$ are  fluctuating, such that their quantum average satisfies: 
 \be   
\brc  j, m  |   J_x^2+ J_y^2 | j,m \ckt = \brc j, m  |    {\mathbbm J}^2 - J_z^2| j,m \ckt =      j(j+1)- m^2 \;.   \label{from2}  
\ee
Each experiment measuring $J_x$, for instance,   observes  one of the possible values,   $ J_x  =  j, j-1, \ldots, -j$, in accordance with
  (\ref{from2}).

Let us check these predictions against Born's rule.  Though  this discussion contains nothing really new,   it may still be refreshing to  see the well-known results 
from a slightly unconventional perspective.

\subsubsection{Spin $\tfrac{1}{2}$}

In the state of spin up, i.e.,   $s_z = +   \tfrac{1}{2}$,
\be   |\!\uparrow\ckt =   |\tfrac{1}{2}, \tfrac{1}{2}\ckt \;,   \label{spinup} 
\ee  
 the measurement of  $s_x$ will give $\pm  \tfrac{1}{2}$,  with no preference for $\pm$   on the average,  and also  with
${\bar  {s_x^2}}  \sim {\bar  {s_y^2}}$,  so we find from (\ref{from2}),
\be  \brc  \tfrac{1}{2}, \tfrac{1}{2} |   s_x^2 |  \tfrac{1}{2}, \tfrac{1}{2}\ckt    =  \frac{1}{4}\;.
\ee
Each experiment will  thus  give the result
$ s_x =   \pm \frac{1}{2},$
with equal relative frequencies   \footnote{ Throughout,  the normalized relative frequencies will be denoted by the calligraphic symbol ${\cal P}$,
while  the normal font letter  $P$  is reserved for the ``probabilities": we hope that the distinction, both formal and conceptual, will not be lost. 
}
\be   {\cal P}_+ =    {\cal P}_- = \frac{1}{2}\;,   
\ee
in accordance with Born's  rule  ($|\pm \ckt$ are the eigenstates of $s_x$):
\be   P_+ =  |\brc + | \uparrow  \ckt|^2 =   \frac{1}{2}\;;\qquad  P_+ =  |\brc - | \uparrow  \ckt|^2 =   \frac{1}{2}\;.
\ee

Consider instead  a measurement of  the spin  component   $s_{n}$   in the generic direction 
\be   s_n\equiv  {\mathbf s} \cdot {\mathbf n}\;, \qquad   {\bf n}  =  (\sin \theta \cos \phi, \sin \theta \sin \phi,   \cos \theta)\;, 
\ee
in the state $  |\!\!\uparrow\ckt  $.   
By rotating the space axes this problem can be refrased as  that of measuring 
 $s_z$   in the state,
\be     | n  \ckt =      c_1 |\uparrow\ckt  + c_2  |\downarrow \ckt  =    \left(\begin{array}{c}e^{-i \phi /2} \cos \tfrac{\theta}{2} \\e^{i \phi/2}  \sin \tfrac{\theta}{2}\end{array}\right) \;, 
\qquad    s_{n}   | n  \ckt =   \frac{1}{2}  | n  \ckt \;,   \label{nspin}
\ee  
i.e., the state in which the spin is directed towards  the unit vector  ${\bf n}$.
The quantum fluctuation of $s_z$  is expressed in the  formulae, 
\be   \brc n | s_z | n \ckt=  \frac{1}{2}  (  |c_1|^2 -   |c_2|^2 )\;, \qquad    \brc n | s_z^2| n \ckt=  \frac{1}{4}  (  |c_1|^2 + |c_2|^2 ) =  \frac{1}{4}\;. 
\ee
Note that this time  the  quantum fluctuations of  $s_z$   towards  $\frac{1}{2}$  and $-\frac{1}{2}$  
are not equally likely, as the spin is directed in a generic direction  ${\bf n}$.  
One finds from the above, 
\be   \frac{1}{2} ({\cal P}_{\uparrow}  -  {\cal P}_{\downarrow} ) =     \frac{1}{2}   (  |c_1|^2 -   |c_2|^2 )\;, \qquad
 \frac{1}{4} ({\cal P}_{\uparrow}  +  {\cal P}_{\downarrow} ) =    \frac{1}{4}\;,
\ee
hence 
\be   {\cal P}_{\uparrow}   = |c_1|^2= \cos^2 \tfrac{\theta}{2}\;, \qquad    {\cal P}_{\downarrow}  =  |c_2|^2=  \sin^2 \tfrac{\theta}{2}\\\;.
\ee
This agrees with Born's rule,
\be  P_{\uparrow} =|\brc \uparrow | n \ckt |^2 =  |c_1|^2\;, \qquad  P_{\downarrow} =|\brc \downarrow | n \ckt |^2 =  |c_2|^2\;.
\ee

\subsubsection{Spin $\frac{3}{2}$} 

Measurement  results of $J_x$  in the state   $|J, J_z\ckt  =  |\tfrac{3}{2}, \tfrac{3}{2}\ckt $  
are dictated by the fluctuation formula  (\ref{from2}),  
\be  \brc \tfrac{3}{2}, \tfrac{3}{2} |   J_x^2+ J_y^2 | \tfrac{3}{2}, \tfrac{3}{2}\ckt =   \frac{15}{4} - \frac{9}{4}  =  \frac{3}{2}  \;.       \label{fromnew}
\ee
This, together with the assumption that the fluctuations in the $x$ and $y$ directions are equal  on the average, and also that  fluctuations towards $J_x >0$ and $J_x <0$ are
 equally likely, yields  
\be    
\frac{9}{4}  {\cal P}(\tfrac{3}{2}) +  \frac{1}{4} {\cal P}(\tfrac{1}{2})  =    \frac{3}{8}\;.  \label{OKagain}
\ee
Actually we have more information on the quantum fluctuations:  we may use  for instance  
\be    \brc \tfrac{3}{2}, \tfrac{3}{2} |   J_x^4 | \tfrac{3}{2}, \tfrac{3}{2}  \ckt =    \frac{21}{16}\;,
\ee
which  gives 
\be     \frac{81}{16}  {\cal P}(\tfrac{3}{2}) +  \frac{1}{16} {\cal P}(\tfrac{1}{2})  =    \frac{21}{32}\;.    \label{OKagainBis}
\ee
Solving the system   (\ref{OKagain}), (\ref{OKagainBis}) we find
\be     {\cal P}(\tfrac{3}{2}) =\frac{1}{8}\;;  \qquad  {\cal P}(\tfrac{1}{2}) =\frac{3}{8}\;.        \label{OKOK}
\ee
On the other hand, 
Born's rule gives  directly
\be     P(\tfrac{3}{2})   = | \brc  J_x=\tfrac{3}{2} |   \tfrac{3}{2}, \tfrac{3}{2} \ckt   |^2   =   
 |  \tfrac{1}{2\sqrt{2}} (1, \sqrt{3}, \sqrt{3}, 1 )\cdot  \left(\begin{array}{c}1 \\0 \\0 \\0\end{array}\right)  |^2= \frac{1}{8}\;;
\ee
\be     P(\tfrac{1}{2})   = | \brc  J_x=\tfrac{1}{2} |   \tfrac{3}{2}, \tfrac{3}{2} \ckt   |^2   =   
 |  \tfrac{\sqrt{3}}{2\sqrt{2}} (-1, \frac{1}{\sqrt{3}}, \frac{1}{\sqrt{3}}, 1 )\cdot  \left(\begin{array}{c}1 \\0 \\0 \\0\end{array}\right)  |^2= \frac{3}{8}\;,   \label{OKagain2}
\ee
in agreement with  (\ref{OKOK}). 
A similar check can be made readily for  the initial state, 
 $|\tfrac{3}{2}, \tfrac{1}{2}\ckt $.

\subsection{General variables} 

The wave function is given by  $|\psi\ckt$, and let us assume that one is now measuring a  dynamical variable $F$.   
We assume that  $|\psi\ckt$ has the form
\be  |\psi\ckt =   \sum_n c_n  | n \ckt\;, \qquad       {F}    | n \ckt = f_n | n \ckt\;, \qquad   \sum_n |c_n|^2   =1\;, \label{stateBis}  
\ee
where    $| n \ckt$'s are  the eigenstates  of the associated self-adjoint operator ${F}$.
For simplicity,  the eigenvalues $f_n$ are assumed to be non degenerate  \footnote{When some $f_n$ are degenerate,  it is necessary to take into account other  variables  $F^{\prime}$ compatible with $F$, and  their expectation values as well. The generalization is, however,  straightforward.    See also  Sec.~\ref{degenerate}. }.

  In the state $|\psi\ckt$  the values of  $F$ are  fluctuating, among all  possible $f_n$'s. 
   The measurement device  each time  picks up one of them,   $F= f_m$, apparently  randomly.     However  the frequency-averaged experimental result 
 \be    {\bar  {f_n}} =      \sum_n  {\cal P}_n  f_n\;,   \label{exactly} 
 \ee
where  ${\cal P}_n$  is  the  expected relative frequency for finding  $F= f_n$,  is equal, by assumption, to the 
quantum fluctuation average (the expectation value),
 \be     {\bar F} =   \brc \psi  |  {F}    | \psi \ckt    =   \sum_n  |c_n|^2  f_n \;. \label{fluctuation}
 \ee
Now,  identification of  (\ref{exactly}) with (\ref{fluctuation}) would be valid   if  
\be       {\cal P}_n  =   |c_n|^2  =    |\brc n |  \psi \ckt |^2\;,    \label{BornG} 
\ee
which, in turn, would follow from  Born's probabilistic rule for  each single experiment, 
\be  P_n =    \brc \psi  |  \Pi_n  | \psi  \ckt =    |c_n|^2      \;, \qquad     \Pi_n = |n \ckt \brc n |\;.  \label{BornG1} 
\ee
In other words,    Born's rule (\ref{BornG1})   is a sufficient condition for  (\ref{exactly}) $\sim$ (\ref{BornG}) to  hold,  but it might in general  not be  necessary.

  In order to show that  it is also necessary,   we must take into account  more information  about the fluctuation of the variable $F$,    
 \be     {\bar O}  \equiv     \brc  \psi | O  | \psi \ckt   =   \sum_n   (f_n)^N     |c_n|^2        \;, \qquad     O= F^N, \qquad N=1,2, \ldots\;, \label{fund1}  
 \ee  
and  identifying them with 
 \be       {\overline  {(f_n)^N}}=   \sum_n   (f_n)^N  {\cal P}_n\;, \qquad N=1,2, \ldots\;.   \label{expect}  
 \ee
 The equivalence of  Born's rule  and  the fluctuation averages  has been illustrated concretely  in  simple spin systems  in the previous section.

In general, if  the eigenvalues $f_n$ are nondegenerate,  and if the Hilbert space is finite-dimensional  (with dimension $D$),  
the expectation values (\ref{fund1})  for $O= F, F^2, \ldots,  F^{D} $ are necessary,  and sufficient,  to yield   Born's rule  (\ref{BornG}).  A simple proof which makes use of the 
nonvanishing Vandermonde determinant is given in Appendix \ref{Vandermonde}.

Actually, a  more straightforward way to see the equivalence between ${\cal P}_n$ and $P_n$, is to consider, instead of various powers of $F$,   a particular function of the variable $F$, which takes the value $1$  if $F= f_n$, and  $0$  if $F \ne  f_n$.     One may construct a Kronecker-delta like operator in terms of the self-adjoint operator $F$, 
\be     \lim_{T\to \infty}    \frac{1}{2T}     \int_{-T}^{T}    d \alpha    \,  e^{ i \alpha (F  - f_n)}  
  \;,  
\ee
but it can be seen to be equivalent  to the 
 projection operator, $\Pi_n  =    |n \ckt \brc n |$.  Its expectation value,   the  expected relative frequency  ${\cal P}_n \times 1 = {\cal P}_n $,      is indeed  equal to 
\be       \brc \psi |   \Pi_n  | \psi \ckt  =      |\brc n | \psi \ckt|^2 =   |c_n|^2\;,    
\ee
but this agrees with Born's rule.  

The discussion can be straightforwardly generalized  to the position measurement, and to  other measurements of any variable with continuous spectrum. In the case of a position measurement, $x$,   the most straightforward way to see the equivalence  is to consider a particular function of $x$,  with small $\epsilon>0$,
\be         f_{x_0}(x) =  \begin{cases}
     \frac{1}{\epsilon}\;,       &        x \in \{x_0, x_0+\epsilon\}\;;      \\
     0\;,  & \text{otherwise}\;.
\end{cases}  \ee
Its expectation value
\be   \brc \psi   |   f_{x_0}  |  \psi\ckt  =  \int dx  \,    f_{x_0}(x)   \, |\psi(x)|^2  \simeq |\psi(x_0)|^2   \;,     
\ee
identified with   ($ {\cal P}(x)$ being the expected relative frequency density) 
 \be     {\bar f}     =   \int dx   \,    f_{x_0}(x)   \,  {\cal P}(x)   \simeq       {\cal P}(x_0)\;,    \label{barfx}  
 \ee
 leads to  (in the limit $\epsilon \to 0$)
 \be      {\cal P}(x) \,  dx   = |\psi(x)|^2   \,dx   \;.\label{position}
 \ee

 \subsection{Summary  \label{SumSec2} } 
 
 The {\it  assumption} that the  expectation value of a general function  $O(F)$ of  an operator  $F$ 
 \be         {\overline  {O(F)}} =   \brc \psi  |  O(F)   | \psi \ckt    =   \sum_n  |c_n|^2  O(f_n)  \; \label{expectSum}
 \ee
is to be interpreted as the prediction for the  experimental average, 
 \be     {\overline  {O({f_n}}}) =      \sum_n  {\cal P}_n \, O(f_n)\;,   \label{exactlySum} 
 \ee
where  ${\cal P}_n$  is  the  (normalized)  relative frequencies  for finding  $F= f_n$,  
leads to 
$  {\cal P}_n =   |\brc n | \psi\ckt |^2 =  |c_n|^2 \;.$

 \section{Pure and mixed states   \label{mixed}} 

A system described by a wave function, (\ref{stateBis})
 \be    |\psi \ckt   =   \sum_n  c_n  | n \ckt   \,,  \qquad  F  | n \ckt = f_n | n \ckt\;,      \label{stateAgain} 
 \ee
which contains in it the complete information on the state,  is known as  a pure state.  
A  pure quantum state contains an enormous amount of information. 
Even a simple spin $\tfrac{1}{2}$ state (\ref{nspin}),  a ``qubit",   can carry a  big quantity of information covering the space, $CP^1 \sim S^2$, to be contrasted with a classical bit, $(1\,\,{\rm or } \, \,   0)$.    We shall use below terms  such as state,  (true) quantum state, genuine quantum state,  and pure state,  
interchangeably \footnote{An exception will be $|\Phi_m\ckt$,  $ |{\tilde \Phi}_m\ckt $, or  $|\Phi_m; Env\ckt$,  $ |{\tilde \Phi}_m; Env\ckt $,    (see Sec.~\ref{measurement}) which describe the experimental device  (and the environment)  after the reading of the measurement result, which we call loosely as ``state"   but  they actually represent  some mixed/classical   state.  }.

If for whatever reason such a complete information is lacking, we have a  so-called mixed state, or a mixture.  A mixed state is described by a density matrix  instead of 
a wave function.  Its definition, the general properties and the way it is used, are well known: they can be found in most textbooks (e.g., Chap. 7 of \cite{KKGP}).  

Here let us just remember a few formulas.   If  the lack of information is due to the fact that the system $A$
observed (and the variables $F$ studied)  is  (refer to)  a subsystem of the total system $\Sigma$,  $A \subset \Sigma$,   the wave function of  $\Sigma$  may be written as
\be    |\Psi \ckt   =   \sum_{n, \alpha}   c_{n, \alpha}   | n  \ckt   |\alpha \ckt    \,,   \qquad    \sum_{n, \alpha}    | c_{n, \alpha} |^2 =1\;,       \label{subsys}
\ee 
where    $\{|n\ckt \} $  is an arbitrary orthonormal set of states   (for instance, the eigenstates of the operator $F$ in  (\ref{stateAgain}))   
describing the subsystem $A$,  and   ${\alpha}$ refer to all variables in $\Sigma/A$.    
The fluctuation average of  a generic variable  $G$  (pertinent to the subsystem $A$)   is encoded in the expectation value,    
\be   
         {\bar  G}  =    \brc \Psi | G   |\Psi \ckt   =      \Tr       {\mathbbm G} \rho \;, \qquad  ( {\mathbbm G} )_{mn} \equiv    \brc m | G | n \ckt\;,  \label{mixedstate}
\ee
where the density matrix is given by
 \be     (\rho)_{nm}  \equiv      \sum_{\alpha}      c_{n, \alpha} \,c_{m, \alpha}^* \;.   \label{density}
\ee
The particular form of the density matrix  (\ref{density})  refers to the case of a subsystem $A \subset \Sigma$.
The trace  formula for ${\bar G}$  in (\ref{mixedstate}), instead, is valid for a general mixed state.    

As said already, a mixed state can arise for any other reason which causes the loss of the complete knowledge on the system.  A well-known example of mixed state is a partially polarized light (i.e., a beam of photons whose polarization states are only statistically partially known).  Another important 
mechanism for the emergence of a mixed state, relevant to our discussion of the measurement processes below, is  that  the measurements, which are  spacetime, pointlike events 
(our  key observation [A]), lead to  the loss of the phase coherence among different terms.     See  (\ref{JumpCor}) and (\ref{diagonal})   below.

The trace  formula (\ref{mixedstate})  generalizes the expression (\ref{fluctuation}) for the fluctuation average in a pure state to the case of a mixed state.

 \section{Factorization vis-\`a-vis entanglement   \label{factorization} }  

   An  issue which has some formal similarity  to the discussions in Sec.~\ref{measurement} below, and is  in fact closely related to them,  is  factorization, which works with a remarkable precision under certain circumstances,  actually making quantum mechanics  a sensible physics theory at all. 
    In simplest imaginable terms, the question is (symbolically)  this: 
   how could we study the physics of a single hydrogen atom, in spite of the fact  that the wave function  must be 
   antisymmetrized with respect to exchanges of {\it  all electrons in the universe}, according to the Fermi-Dirac statistics for identical
   fermions?  
   
    The answer is that, for instance for the two electrons, one in the laboratory, the other on the sun,  the wave function is given by 
   \be     \Psi = \frac{1}{\sqrt{2}}    \left(  \psi_{lab}({\mathbf r}_1)  \psi_{sun}({\mathbf r}_2) -     \psi_{lab}({\mathbf r}_2)  \psi_{sun}({\mathbf r}_1)  \right)\;, \label{sun} 
   \ee
   but the second term is utterly negligible when   
   \be      {\mathbf r}_1 \in {\rm  Lab.} \;, \qquad    {\mathbf r}_2 \in {\rm Sun}\;,      
   \ee
   as the wave function $ \psi_{lab}$   (vis-\`a-vis,  $\psi_{sun}$)  has (spatial)  support in a small region in 
   the laboratory (in the sun).  There is no overlap. Only the first term in (\ref{sun}) survives, and  the laboratory electron is simply described by the wave function   $ \psi_{lab}({\mathbf r}_1)$.  
     
Instead of this symbolic example,  one may consider the second hydrogen atom 
   in the next room in the laboratory, e.g., two meters away.  From  Bohr's radius for the hydrogen atom, one finds  a suppression of the order of  
   $10^{-18}$ for factorization failure.    This rule-of-thumb estimate may give some idea about the exactness of our assertion in  Sec.~\ref{measurement}, even though 
   there the question will be  the lack of the overlap in {\it spacetime} supports.

     Just to close a loophole in the discussion, we note that   if the antisymmetrization is done with respect to the spin state, and not with
     respect to the space positions as in (\ref{sun}), the wave function does not  factorize  (the two electrons are entangled).    In that case, we have  an unknown mixture for the spin state
     of the laboratory electron.   
     
    In fact, factorization and entanglement are two faces of the same medal, both characterizing quantum mechanics universally.  
     Certain aspects of quantum entanglement 
  such as those  briefly reviewed and discussed  below in Sec.~\ref{entanglement} in connection with the EPR paradox,  have become one of the most ardently discussed topics of quantum mechanics, presumably  because of some fascinating features involved  (e.g., quantum nonlocality). 
  
  Actually, the occurrence of entanglement is much more general and indeed, ubiquitous. It does not depend on the Fermi-Dirac or Bose-Einstein statistics either, contrarily  to what  the above example might have erroneously  suggested.  All systems which have interacted  with each other in the past are  generally  entangled.  
  Any quantum system in which factorization does not hold in good approximation, is entangled   and is, typically,  in a mixed state.
  Our  whole world is a large quantum system, in a complex, entangled mixed state. Macroscopic bodies  in it  generally exhibit classical behaviors because of entanglement with their environment \cite{Zurek,Joos}.

   It is the factorized quantum states -  those described by the wave function, e.g.,  of an atom or  of an electron  -   {\it   that are truly extraordinary, and exceptional.  }
   The concept of a pure state  $|\psi\ckt $   depends on the factorization of the system described by $\psi$,  from the rest of the world.    These states  are carefully prepared by the experimentalist, inside many small  isolated bubbles in the world (i.e., physics laboratories), by using sophisticated vacuum and clean-room  technologies.   Pure states (free particles)
   occur also in Nature,  as a result of radioactivity on Earth, or as cosmic rays in interstellar space, for instance.
   See Sec.~\ref{Universe} for more discussions.

  As will be seen below,   
 factorization of $|\psi\ckt$  describing the microsystem which is the object of the measurement  gets rapidly ruined during the measurement process    as a result of the interactions  between it and
  the measurement device as well as  with the rest of the world,  and  a
 particular  set of entangled mixed states  labelled by  the measurement readings  emerge. 
  The result is the  ``wave-function collapse", or better,   what may be perceived as such,  see the discussions below.

  \section{Measurement processes \label{measurement} }  
 
 To have  a complete picture of the measurement processes  it is often assumed  that the full wave function contains the microscopic
 state $|\psi\ckt$, the apparatus state  $|\Phi\ckt$,  and  the state of the environment,  $| Env \ckt$, the last containing everything outside the experimental device,  such as the air molecules,  the computer screen, the experimentalists, etc.  The measurement process, on the state  \footnote{For definiteness, and in order not to introduce 
 other   (though interesting) issues, we assume  here that $F$, as well as the Hamiltonian of the system, are  time-independent operators.   }
 \be    |\psi \ckt   =   \sum_n  c_n  | n \ckt   \,,  \qquad  F  | n \ckt = f_n | n \ckt\;,      \label{stateBisbis}   
 \ee 
 is supposed to proceed as   (the states with index $0$  stand for  the neutral state, before the reading of the results):
 \bea     |\psi \ckt \otimes |\Phi_0\ckt  \otimes   | Env_0 \ckt  &=&   \left( \sum_n  c_n  | n \ckt  \right)    \otimes |\Phi_0\ckt  \otimes    | Env_0 \ckt   \label{step0}  \\
   & \longrightarrow &      \left( \sum_n  c_n  | n \ckt     \otimes |\Phi_n\ckt  \right )   \otimes   | Env_0 \ckt      \label{step1} \\
  & \longrightarrow &     \sum_n  c_n  | n \ckt     \otimes |\Phi_n\ckt    \otimes   | Env_n \ckt   \;,\label{step2} 
 \eea
 where in the first step  (\ref{step1})  the experimental device has read the measurement results,  $f_n$, and in  the final step   (\ref{step2})  the environment has come to be aware of such a result (the experimentalist has seen  the 
 result on her/his  computer screen).  

  This kind of formulas, apparently describing coherent superpositions of distinct  macroscopic states,  has given rise to innumerable debates 
 in the literature  as, most famously,  in the so-called Schr\"odinger cat conundrum. 
 Actually,  the Schr\"odinger's cat  discussion involves several extra issues which are not present in the general measurement processes
considered here.
 To avoid unnecessary complications and  unavoidable confusion,  we  will postpone its discussion to Sec.~\ref{cat}. 
 
 Actually,  the factorized form of the state vectors in (\ref{step0}) $\sim$ (\ref{step2})  (various $\otimes$ symbols), found often in the literature,  is not valid in general,   
 except for the first. The factorized form of the wave function  of the microsystem  $|\psi\ckt$   before the measurement,  in the first line,  (\ref{step0}),
 is correct by definition: we are assuming to have a microsystem described by $|\psi\ckt$  before the measurement.

    The  role of the environment  or the entire world outside the measurement device,   $|Env\ckt$, in measurement processes,  is subtle. 
    The factorized form $|\Phi\ckt \otimes |Env\ckt$  is not in general guaranteed,   even before the actual experiment,  i.e.,  even in (\ref{step0}).  
     It suffices to consider the air molecules around, the casing of the device itself, etc., which all represent complicated entanglement between $|\Phi\ckt$  and $|Env\ckt$. This means also that the boundary (or distinction) between what is  to be considered as the experimental device and what belongs to the rest, is not well defined. More precisely, the exact boundary between the two is largely arbitrary, and must be regarded as conventional. 
     
Also,   the experimental device,   after the reading of the measurement, is in a classical state.  This fact,  the emergency of the classical behavior of the experimental apparatus, is believed to be due to the entanglement  \cite{Zurek,Joos}   between $|\Phi\ckt $ and $|Env\ckt$.   
   
In spite of all this,    every experimentalist  knows perfectly well  
which part of her (or his)  device is essential, and that the blurred boundary between it and  the environment  does not affect her (his)  results,  to any significant degree.   She (or he)  would not worry  where  the exact boundary between the measurement device   $\Phi$   and the environment   lies,
   and certainly would not play with the idea of  pushing  this boundary up to inside the human brain, as has been sometimes done by theoretical physicists.  She (or he) knows that there is no need to do so, in order to assure good, precise experimental results. 
   
  Having all these subtleties and understanding in mind,  we  will write the  measurement device-environment ``state"  as   
   \be       | \Phi;  Env  \ckt 
   \ee
   below,    instead of    $|\Phi\ckt \otimes |Env\ckt$.      
    The general measurement process can be described  as   \footnote{The device-enviroment ``state", $ |\Phi_0;    Env_0 \ckt $,  
    is well defined at each $t$, before any measurement. 
    However it  can never be identical,  at  two different measurement instants,  see Sec.~\ref{loss} below.  }
       \bea     |\psi \ckt \otimes |\Phi_0;  Env_0 \ckt  &=&   \left( \sum_n  c_n  | n \ckt  \right)    \otimes   |\Phi_0;    Env_0 \ckt   \label{step000}  \\
   & \longrightarrow &       \sum_n  c_n  \,    |{\tilde  \Phi}_n;     Env_0 \ckt      \label{step111} \\
  & \longrightarrow &     \sum_n  c_n   \,   |{\tilde  \Phi}_n;     Env_n \ckt       \;.\label{step222} 
 \eea
 where  $ {\tilde  \Phi}_n $ stands for  the entangled state of the microsystem-apparatus with the reading of the measurement result, $f_n$.  The exact timing 
 of passage from  the first stage of  measurement-registering of the result on the apparatus (\ref{step111})  to the  second (\ref{step222})   (e.g.,  the moment  in which  the experimentalist has seen the  
result  on her  or his computer screen) will depend on the largely arbitrary division between   $\Phi$ and $Env$;  but it will not be important.

 In special, ``repeatable" experiments   the process  may take a special form,   see  Sec.~\ref{WFcollapse} below,
  in which   the result of each  measurement can be used as a  state preparation.

 \subsection{Emergence of  mixed states and state-vector reduction \label{loss}} 
 
\begin{figure}
\begin{center}
\includegraphics[width=7.5 in]{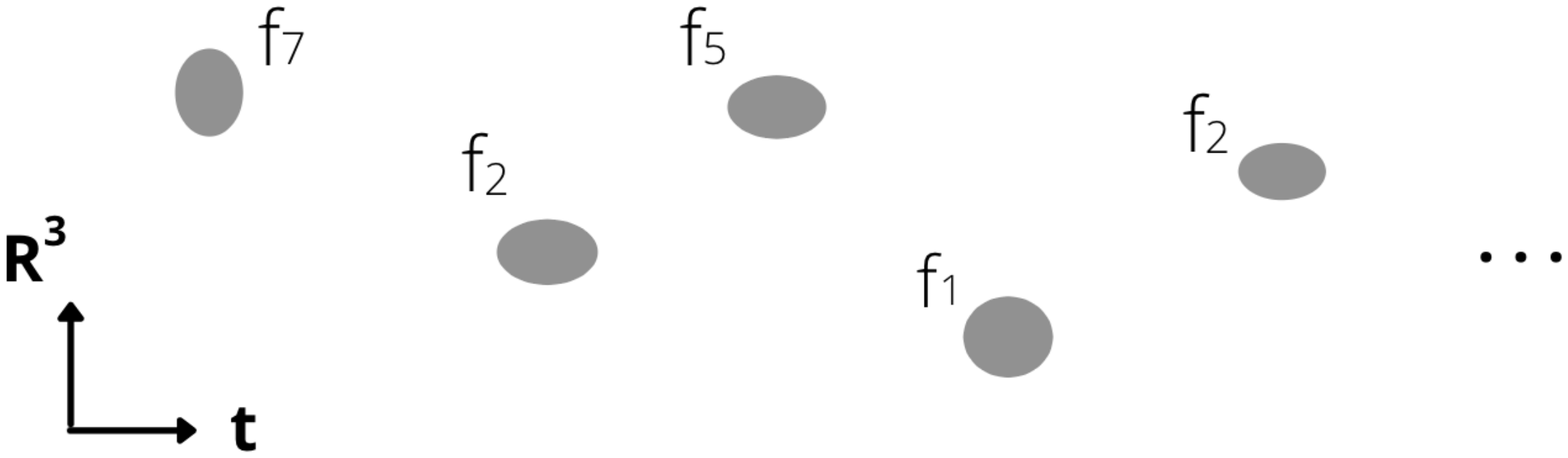}
\caption{\footnotesize Measurement of a variable $F$.  Each blob describes a single measurement event (a chain ionization reaction, hadronic cascade, etc.) occurring at a localized region in spacetime, 
with the experimental results,  $F=f_7, f_2, f_5, f_1, f_2$, etc.  The wave functions describing the  different measurement events ${\tilde  \Phi}$ have no overlapping spacetime supports, therefore are orthogonal.
}
\label{mixture}
\end{center}
\end{figure}

The way out of the macroscopic interference  effects  and consequent paradoxical features,   comes from 
  the key ideas in this work,    that   [A]  each measurement is  a space-time pointlike  event,  entangling  [B]  the microsystem with the experimental device-environment
  ``state".        The experiment which gives   rise to one result 
$F= f_m$  and another with the result  $F= f_{\ell}$,  are  two distinct spacetime events.
  The   wave functions   $  |{\tilde  \Phi}_m;     Env_0 \ckt   $ and  $  |{\tilde  \Phi}_{\ell};     Env_0 \ckt $  associated with two distinct measurements 
  (including the case  $m =  \ell$),
{\it   have spacetime supports which do not overlap,}  see Fig.~\ref{mixture}. There are no  interferences among different terms  in (\ref{step111}).

Before the measurement takes place, (\ref{step000}),   the microsystem   is described by the wave function,  
\be  |\psi\ckt=  \sum_n c_n |n\ckt\;,  \qquad    F | n \ckt = f_n | n \ckt\;:  \label{thestate} 
\ee
it is a pure state.   It is actually  a subsystem of the whole  world  (as in  (\ref{subsys})),    but, as we have already noted,  it  is described by a wave function of factorized form by assumption, 
\be    |\Psi\ckt_{t=0_-}  =      \left( \sum_n  c_n  | n \ckt  \right)    \otimes   |\Phi_0;    Env_0 \ckt  \;.
\ee
The expectation value of any  generic  variable  $G$ in  this state (before the measurement) is given by  
\be   {\bar G}=    \brc \Psi | G  | \Psi\ckt   =       \Tr       {\mathbbm G} \rho \;, \qquad  ( {\mathbbm G} )_{mn} \equiv    \brc m | G | n  \ckt\;, 
\ee
with the  density matrix  having the form  
\be  \rho=    \left(\begin{array}{ccccc}|c_1|^2  &  c_1 c_2*   &  c_1 c_3^*  &   \ldots  &  \ldots  \\   c_2 c_1^*    & |c_2|^2 &  c_2 c_3^* & \ldots   &  \\   
c_3 c_1^*  &  & \ddots &  &  \\  \vdots    &  &  & |c_n|^2 &  \\ &  &  &  & \ddots\end{array}\right)\;,   \label{pure}
\ee 
characteristic of a pure state, i.e.,   $\rho_{nm} =  c_n c_m^*$.    The normalization condition 
\be          \brc  \Phi_0;    Env_0   | \Phi_0;    Env_0 \ckt  =1\;
\ee
has been used. 
In the case of the particular variable  $F$  (whose eigenstates are $|n\ckt$'s),  $  {\mathbbm F} $  is diagonal,  and its  quantum average is given by  the known formula  (\ref{fluctuation}),  
\be   {\bar F} =    \Tr  ( {\mathbbm F} \rho )  = \sum_n   f_n  |c_n|^2\;. \label{usual}  
   \ee
   
As soon as the microsystem gets into contact with the experimental device, and the measurement events (a chain ionization process, hadronic cascade, etc.) have taken place,  the total wave function takes an entangled form, (\ref{step111})  \footnote{$ {\tilde  \Phi}_n $ is the entangled microsystem-apparatus  state with the reading of the measurement result, $f_n$.  },
\be    |\Psi\ckt_{t=0_+}  =     \sum_n  c_n      |{\tilde  \Phi}_n;     Env_0 \ckt  \;.
\ee
Now, the wave function describing   ${\tilde  \Phi}_m$  (the aftermath of a measurement, with the result, $F=f_m$) and  that for    ${\tilde  \Phi}_{n}$
 (the aftermath of a measurement, with the result, $F=f_{n}$),  corresponding to two  distinct spacetime events,   have no overlapping 
 spacetime support, as illustrated in Fig.~\ref{mixture}.  
   The  expectation values of a variable  $G$ are now  given by
   \be   {\bar G}=    \brc \Psi | G  | \Psi\ckt   =       \sum_{m,n}  c_m^* c_n  \brc  {\tilde  \Phi}_m;     Env_0  | \, G \,  |   {\tilde  \Phi}_n;     Env_0 \ckt   \;,
   \ee
   but the fact    ${\tilde  \Phi}_m$  and ${\tilde  \Phi}_{n}$   ($m \ne n$)     have no common spacetime support,
     means that  {\it   for any local operator } $G$,   orthogonality relations, 
       \be      \brc    {\tilde  \Phi}_m;     Env_0     |   \, G \, |     {\tilde  \Phi}_{n};     Env_0 \ckt =0\;, \qquad  m \ne n  \;, \label{lack}  
\ee
   hold.     Consequently,   $  {\bar G} $ is given by  the sum of the diagonal terms
   \be    {\bar G} =    \sum_n   |c_n|^2  G_{nn} \;,    \qquad   G_{nn}=    \brc    {\tilde  \Phi}_n;     Env_0     |  \,G \,|    {\tilde  \Phi}_{n};     Env_0 \ckt \;.     \label{above} 
   \ee  
This     means  that the density matrix has been  reduced to a  diagonal form,   
\be  \rho=    \left(\begin{array}{ccccc}|c_1|^2  &     &    &  &  \\    & |c_2|^2 &  &  &  \\      &  & \ddots &  &  \\  &  &  & |c_n|^2 &  \\ &  &  &  & \ddots\end{array}\right)\;.\label{diagonal}
\ee 
For the particular case of the variable  $F$,   we recover the standard prediction, 
\be    {\bar F} =    \sum_n   |c_n|^2   f_n \;, \qquad  .^.. \quad    {\cal P}_n=    |c_n|^2\;, \label{recover} 
\ee
where $ {\cal P}_n$ are the normalized relative frequencies for different outcomes  $f_n$.
    
   The emergence of the mixed state   (\ref{diagonal})   \footnote{Taking into account the fact  that the microsystem-apparatus combined system after the measurement is a mixed state,  no longer a pure state, eliminates the contradiction discussed in a recent paper \cite{Renner}. Another possible issue in \cite{Renner}  seems to be 
the fact that the information transmission from the first laboratory to the second  about the result of the measurement made in the first, renders invalid the assumption that the two laboratories are isolated
quantum systems. 
}  is generally attributed to the fact that  a typical  experimental device is made of a macroscopic  (difficult-to-specify)  number of atoms and molecules, and it is not possible to keep track of the phase relations among different terms  in (\ref{step111}) to any significant extent.  Also,  the macroscopic device  $\Phi$, evolving in $t$ entangled with $Env$, can never be in an identical quantum state 
$ |\Phi_0; Env_0 \ckt $   at  two different  measurement instants.  This is in stark contrast with  the  {\it  identical quantum state } of the microsystem, $|\psi\ckt$, which can be  and is indeed produced via e.g., a repeatable experiment (state preparation - see Sec.~\ref{WFcollapse})    for each  measurement \footnote{The fact that  atoms (e.g., the hydrogen) of the same kind in the  gound state, are all   in a  rigorously identical quantum state,  is at the base of the regular structure of our macroscopic world. }.

The observations above are certainly correct, but  we need also another,
 crucial ingredient for  the decoherence in the measurement processes:   the lack  of the common  spacetime supports  in the wave functions,  and the consequent orthogonality of terms corresponding to different measurement terms,  (\ref{lack}).  Importance of this is that  it implies  that 
 the result of each measurement event  is a state-vector reduction,   
\be    \left( \sum_n  c_n  | n \ckt  \right)    \otimes |\Phi_0; Env \ckt    \Longrightarrow     |{\tilde \Phi}_m ; Env \ckt \;:      \label{JumpCor}
 \ee
 i.e.,  with a single term present,  the instant after the measurement (e.g., with $F=f_m$).    
 This state of affair is perceived by us as a  ``wave-function collapse".  
 
 Summarizing:
 \begin{description}
  \item[A:]    The spacetime eventlike nature of the triggering  particle-measuring-device interactions  introduces an effective spacetime localization of each measurement event; 
  \item[B:]   At the moment  the microsystem - a pure quantum state $|\psi\ckt$ - gets into contact with the measurement device-environment state $ |\Phi_0; Env_0 \ckt $  which is a mixed state,
     factorization of $|\psi\ckt$ gets lost rapidly, and an entangled, mixed (classical) state   $ |\Phi_n; Env_0 \ckt$ with the unique recording,  $F= f_n$,  is generated.  
\end{description}
The result is the state-vector reduction,  (\ref{JumpCor}).

\subsubsection{Particle hitting a potential barrier  \label{barrier}} 

To clearly grasp the idea of effective spacetime  localization of the measurement events,  
 consider a simple particle scattering off a one-dimensional square 
 potential barrier.  Even though  the plane-wave description yields the correct transmission  ($D$)  and reflection   ($R$)  coefficients, 
as taught  in every quantum mechanics course,  a time-dependent, {\it  wave-packet}  description of the process shows that the initial rightmoving incoming packet,  the  transmitted wave packet travelling rightwards
and the reflected wave packet  travelling backward towards left,    do not overlap with each other in spacetime. The absence of the interferences between any two of them
is thus seen trivially in the wave-packet  description,  while it is 
not  obvious 
in the static,  plane-wave formulation of the problem commonly employed.   To complete the analogy with 
(\ref{step111}), one may consider measuring the particle at both sides of the barrier. Each time, the particle
will appear  {\it   either} on the right  {\it or }  on the left, with average relative frequencies,   $D$ and $R$, respectively.

\subsubsection{Tonomura's  ``double-slit" experiment \label{double}}

The  experiment by Tonomura et. al. \cite{Tonomura}     is a perfectly general  type of  the measurement process, described by   
(\ref{step000}) $\sim$ (\ref{step222}).  Before the electron hits the photographic plate,   it is described by a wave function 
$\psi$   having two components coming through the two sides of the filament with positive potential between two plates  (acting as a double slit \cite{Tonomura}).
The measurement takes place at the moment the electron impinges on the photographic screen,  and leaves an ionization blot at  some $x$.

\begin{figure}
\begin{center}
\includegraphics[width=2.4in]{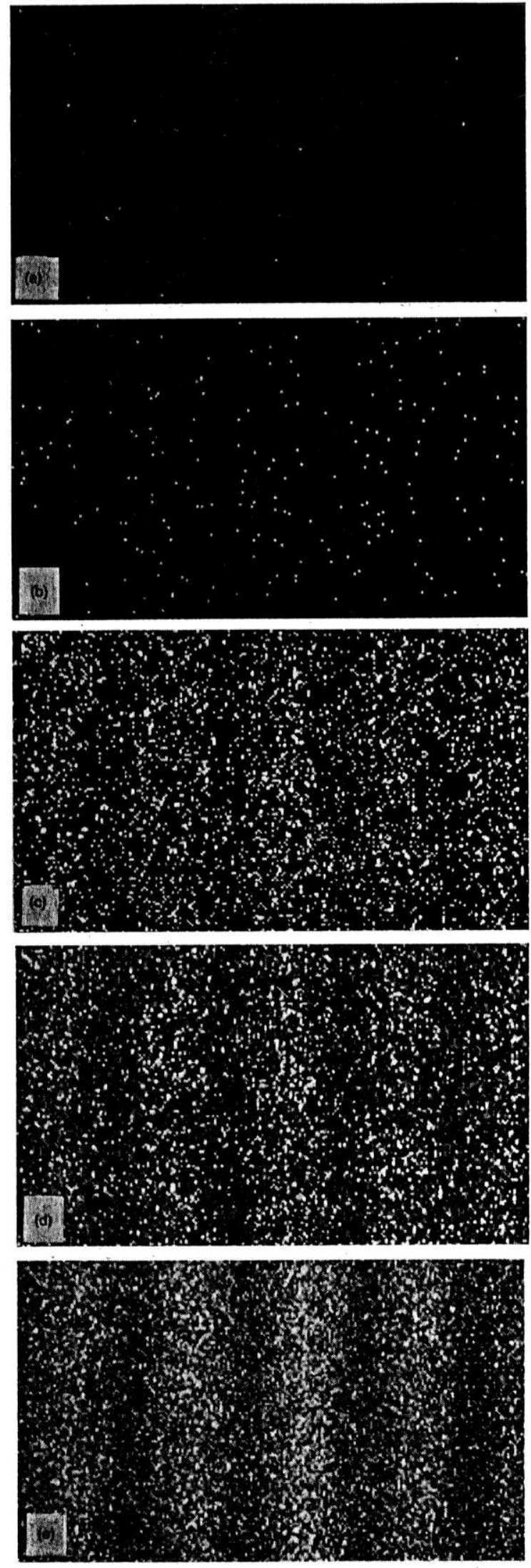}
\caption{\small  The pictures correspond, from the top to the bottom,  to successive  exposures to $10$, $100$, $3000$,  $20000$ and $70000$ electrons, respectively.
Reproduced from   \cite{Tonomura},    A~. Tonomura, J~. Endo, T~. Matsuda, T~. Kawasaki and H.~Ezawa,
``Dimonstration of single-electron buildup of interference pattern",
  American Journal of Physics  57, 117 (1989); doi: 10.1119/1.16104, 
 with the permission of the American Association of Physics Teachers.
}
\label{TonomuraExp}
\end{center}
\end{figure}

  In this experiment
 the electron beam is carefully prepared.
 The electron wave packets, the mean velocity of the electrons and the beam intensity, are such that  the average  distance between the two successive electrons are  (so to speak)  $\sim150\, km$, to be compared with the size of the wave packets ($\sim 1 \, \mu m$)   and the distance from the source to the screen   ($\sim 1.5 \, m$).  The wave packets of different electrons do not interfere with each other. The electrons do  arrive,  one by one    (Fig. \ref{TonomuraExp}).

  The normalized expected frequency for different values of $x$ is given by   $ \sim |\psi(x)|^2$,  which predicts interference fringes \`a la Young.  They are indeed observed experimentally,  after many electron images  ($20000\sim70000$) are  collected, see Fig. \ref{TonomuraExp}.
  
  Note that different points on the screen   (ionization blots)  correspond to different spacetime events (different electron arrivals, different   measurements): the wave packets of different electrons do not overlap in spacetime.   Each ionization blot in Fig. \ref{TonomuraExp}  represents a measurement blob of  Fig.  \ref{mixture},   but here everything is real physics, not a caricature.

  In both examples  of Sec.~\ref{barrier} and Sec.~\ref{double},    {\it    the very possibility of considering, or preparing experimentally,  wave-packets of arbitrarily small size, hinges upon  the particle nature of the degrees of freedom of our interest. Arbitrarily small  wave packets  mean that the approximation of neglecting the  coherent superposition among different terms in (\ref{step111})   can be made  as good as we wish.  }

   \subsection{Repeatable,  nonrepeatable  and intermediate types of  state reductions
   \label{WFcollapse} }

    In an exceptional class of  (``repeatable") experiments,  the  processes  (\ref{step000}) $\sim$ (\ref{step222})  may look instead  as,
     \bea     |\psi \ckt \otimes |\Phi_0;  Env_0 \ckt  &=&   \left( \sum_n  c_n  | n \ckt  \right)    \otimes   |\Phi_0;    Env_0 \ckt   \label{step00000}  \\
   & \longrightarrow &       \sum_n  c_n     |n\ckt \otimes   |{\Phi}_n;     Env_0 \ckt      \label{step11111} \\
  & \longrightarrow &     \sum_n  c_n      |n\ckt \otimes   |{\Phi}_n;     Env_n \ckt       \;:   \label{step22222} 
 \eea
namely, each of  the states of the original microsystem $| n \ckt$ remains factorized from the apparatus-environment state,  
 even though the latter has registered the  reading  $F= f_n$. 
 The state vector reduction (\ref{JumpCor})   takes a special form,   
\be    \left( \sum_n  c_n  | n \ckt  \right)    \otimes |\Phi_0; Env \ckt    \Longrightarrow       | m \ckt  \otimes  |{\Phi}_m ; Env \ckt \;.   \label{JumpRep}
 \ee
Namely the measurement leaves  the microsystem with a well-defined wave function, $|m\ckt$.  
This kind of processes  is  often discussed in textbooks,  as a typical ``wave function collapse", 
\be    |\psi\ckt =     \sum_n c_n  | n \ckt   \Longrightarrow   | m \ckt\;. \label{collapse}
\ee
A  process such as  (\ref{JumpRep}) or (\ref{collapse})  means  that  the result of an experiment can be used as the initial condition for 
 subsequent studies of the system,  i.e.,  as a state preparation. This  (ideal) type of measurements are known as ``repeatable" in the literature.

Even though  these represent indeed special class of measurements, they are neither rare nor particularly difficult to realize experimentally, either.  More importantly, they play an essential role for the consistency of  quantum mechanics (see  Sec.~\ref{loss} and  Appendix \ref{subtle}).
 
  An almost trivial example  of a repeatable measurement  is a position determination by a little sheet with a tiny hole. If a particle has passed it, its position has been measured. 
Let us superpose another sheet with a hole at the same position. The particle which passes the first hole passes also the second.  But now overlay it with a second sheet with a hole in a different  position.   The particle which passed the first gets blocked by the second, this time. This over-simplified model describes  nonetheless a perfectly valid quantum mechanical measurement, illustrating (\ref{collapse})
\footnote{This example was inspired  by the electron wave ripples obtained by  A. Tonomura, by letting $50$ kV electron beams go through a collodion thin film with tiny holes. The  images, which look identical to  water ripples  on a puddle surface  produced  by rain drops,  but smaller by a factor 
$\sim 10^{-6}$ in size,  can be seen, enlarged appropriately,  on the front-cover page of the book \cite{KKGP}.
}.

Another,  more  interesting example of the repeatable measurement,  is a variation of the standard  Stern-Gerlach (SG) experiment with an inhomogeneous magnetic field with gradient in the $z$ direction, with a beam of massive spin $\tfrac{1}{2}$  atoms moving in the $x$ direction.  
 
 Let us assume for definiteness  that the spin of the atoms is directed towards ${\bf n}$:    it is in the state (\ref{nspin}).   
 As the atoms proceed, their wave packets are divided in two, one going up (spin $ s_z=\tfrac{1}{2}$) and the other  going down   (spin $ s_z=-\tfrac{1}{2}$)
  \footnote{ In view of the discussions on  macroscopic quantum systems later, Sec.~\ref{MacroQM},  we note a very interesting aspect of the standard SG experiment. It is the fact that,  unless the final position measurement of the atom  is done, the two wave packets maintain the phase coherence,   even if they are separated by a macroscopic distance
and have momentarily lost common spatial support.  They are still described by the single wave function of the atom: it is a sort of  macroscopic quantum-mechanical state
(Sec.~\ref{MacroQM}, and Sec.~\ref{entanglement}). Their interference effects can be observed, once  the two wave packets are carefully  re-converged  with an appropriate second magnetic field, in an experimental arrangement  known as the  ``quantum eraser".}.

 Upon arrival at the photographic plate the atoms will leave two bands of ionization blots, upper and lower ones, with relative intensities, $|c_1|^2$ and  $|c_2|^2$.   
 This standard  SG experiment is  clearly  of a generic  (not repeatable) type,   described by   (\ref{step000})$\sim$  (\ref{step222}).

Let us now, however,  place a concrete block to  impede the lower  wave packet from proceeding further towards right. 
On the  photographic screen at the end,  we will observe  just one  (upper) band of the images, instead of  two.  This measurement is still not repeatable.

But now remove the photographic plate (but not the magnetic field and the concrete block). This time apparently nothing will be  observed:  the measurement is not completed.

 Nevertheless, the beam of the atoms which have passed the region of the magnetic field and cement block,    
contains now only  atoms in the spin-up state.  Related processes are known as
``null measurements" in the literature  \footnote{Other terms such as "interaction-free experiment" or related    "negative-results experiment" 
have also been used in the literature, but the idea is similar.};   we may coin them as "half measurements"  as well, as only the  first half  of the experiment  - the preparation of the magnetic field and of the cement block has been done, but no final observation of the  vertical position of the atoms in arrival  on the screen-  has been made.  

But because of this,  the part of the beam which passed the experimental region  contains only atoms in the state $ s_z=\tfrac{1}{2}$,  and can be used in subsequent experiments  with the known initial wave function,   $|\!\uparrow \ckt $.    The state-vector reduction has occurred  exactly as in (\ref{JumpRep}), (\ref{collapse}). 
 Note that 
for subsequent studies only the atoms which passed the first experimental region matter.  

 Still another example of a repeatable measurement  is a beam  of  light  impinging on a crystal or a polaroid,  which have a given polarization axis,  
discussed in many textbooks.

However,    in more general types of experiments this aspect  (the measurement as the state preparation)  cannot be maintained.  
Often the measurement takes place when an incoming electron, photon, proton, atom,  etc. triggers  a chain ionization process or a hadronic  cascade 
 at some point of the measuring device,  leaving a particle track on the instrument (registered on time), or an ionization blot on the photographic plate. It might look as if what ``collapses"
 is not the wave function of the incoming electron  but  the equilibrium state of the macroscopic matter which make the body of the device, from a neutral, uniform state   to the one 
 with a particular reading of the measurement result.    The incoming electron gets simply  lost somewhere.  
  The state-vector reduction takes the general  form  (\ref{JumpCor}): 
 the measurement does not serve as a state preparation
  for subsequent studies.

For completeness,  
  let us note that   the state reduction  of  {\it repeatable} experiments,  (\ref{collapse}), (\ref{JumpRep})    where the result of the measurement serves as the preparation of a well-defined quantum state,   and  general {\it non-repeatable}  processes,  (\ref{JumpCor}), in which the original microsystem 
(or information about it) gets completely lost after the measurement,      represent  actually  only   two extremal situations.   There exist  many intermediate types of experiments  in which  the information about the original microsystem is  only partially maintained (or  partially lost), during  the measurement \footnote{Expressions such as ``partial wave-function collapse" have been also used in the literature.}.  The $\alpha$-particle track in a Wilson chamber, the hadronic cascade in a calorimeter, the particle tracks in a silicon detector,   are
all examples of  these more general types of state reductions.  The momentum or  energy measurement in high-energy experiments,  typically requires not a single spacetime event but a sequence of related sub-events.     

We will not make, however,    any attempt in this work  to formulate and classify these different kinds of measurement processes.   The concept of the state-vector reduction as  used in Sec.~\ref{measurement},   
is valid and applies  equally well to all these cases.

   \subsection{Unitarity  and linearity   \label{unitarity}  }   
      
   The microscopic state  $\psi$   
\be      |\psi\ckt = \sum_n c_n  | n \ckt\;,      \label{stateU} 
   \ee
  if left undisturbed,  evolves in time according to the Schr\"odinger equation, 
  \be  i \hbar  \frac{ d }{d t}   |\psi\ckt    = H\,  |\psi\ckt  \;, \label{Schr}
  \ee
 ($H$ being the Hamiltonian of the system)   or as
   \be      |\psi\ckt    \longrightarrow    U\,   |\psi\ckt =     \sum_n   c_n  \,U  | n \ckt \;,    \qquad  U= e^{- i H t / \hbar}\;,    \quad    U^{\dagger}  U = {\mathbf 1}\;:   \label{Hin}
   \ee
   it is a linear, unitary evolution.   
   Unitarity  means that
   \be    \brc  \psi | \psi \ckt =   \sum_n  |c_n|^2  =1\;,  \label{normaU} 
   \ee
   i.e.,  the state is normalized and its  norm is conserved in time.    Linearity of evolution  means  (\ref{Hin}), i.e.,  superposition of different states  in (\ref{stateU})   continue to be  coherent superposition of the corresponding states,  
   though each term has evolved by the same unitary operator $U$. These  properties of  pure states  are  well-known.

 During the measurement process,  when the microsystem gets  in contact with the measuring device, the latter reading the 
   result of the measurement,  the time evolution, formally written as   (\ref{step111}), might  still look unitary and linear.   
   However, as we have seen,  the wave functions associated with different terms in   (\ref{step111})  do not overlap in spacetime:   it represents a mixed state.  
   The coherent superposition of states is no longer there, see (\ref{lack}). 
   Most significantly,  the real time  evolution 
   of the system is the state-vector reduction,   (\ref{JumpCor}).

   This means that  linearity is lost during the measurement.

   On the other hand,  unitarity  is maintained \footnote{Note that linearity and unitarity  are two distinct concepts in quantum mechanics  (e.g., the momentum operator is linear but not unitary).    Dirac    noted   (Chap 27 of \cite{Dirac})  that the evolution operator must satisfy both, which are two independent requirements. They are both automatically met, once the choice  $U= e^{- i H t/\hbar}$ is made with a self-adjoint Hamiltonian operator,   and as long as the pure-state evolution  (\ref{Hin})   {\it  before}  the measurement,   is concerned.    
   }: 
   the sum of all possible outcomes, occurring at different measurements at different times,  adds up  to unity of  the total normalized frequencies,  corresponding to the  ``norm"   of the   state,  (\ref{step111}).
   In terms of the relative frequencies for different outcomes,    ${\cal P}_n$,    unitarity reads
   \be    \sum_n   {\cal P}_n  = 1\;,    \qquad \sum_n  |c_n|^2 =1\;,  \label{obvious} 
\ee
even though different (sometimes the same, but repeated) experimental results $f_n$  refer to distinct measurements made at different times.

   Note that   this is  indeed  how experimentalists  view the meaning of unitarity.   Unitarity means that   $ {\cal N}_n / {\cal N}$  \footnote{ ${\cal N}_n$ is the number of times the experimentalist finds the result  $F=f_n$,  $n=1,2,\ldots$;   ${\cal N}$ is the total number of the measurements made.}, to be identified with the theoretical formula,  $   {\cal P}_n=    |c_n|^2$,
   should satisfy 
   \be \sum_n  {\cal N}_n / {\cal N} =1\;,  \label{trivial2}   \ee   
     which might look trivial.  However,  this contains an implicit, important   assumption  that 
 the  experimental device has no systematic bias, and registers all possible results $f_n$ with equal efficiency and with no losses.   
 Only in such an ideal measurement setting we can expect  that the experimental results $ {\cal N}_n / {\cal N}$ will approach  
  the  theoretical  prediction, ${\cal P}_n$,   {\it  at large ${\cal N}$}  (e.g., Tonomura's experiment, Fig.~\ref{TonomuraExp}).

  It is interesting to compare the concept of unitarity in the measurement processes  as described here, with 
a more abstract  one,  in the conventional thinking based on Born's rule.   The latter means in fact
\be  \sum_n      P_n  = 1\;,\label{SumP}  
\ee
i.e., that the total    {\it probability}  for a single experiment   to give all possible results $f_n$,   is unity, and that it is conserved in time. 
The request  (\ref{SumP}), from the mathematical, logical point of view,  appears  quite  indispensable, and indeed has always been considered sacrosanct in the traditional approach to quantum mechanics.

However,  once   $P_n$  is translated into the physically directly meaningful quantity, 
 ${\cal P}_n  ~(=P_n) $,
  the expected relative frequencies for various results $f_n$ which will  be 
 found in distinct measurements and  at different times,  unitarity as 
 an absolute principle of the conservation of the total probability,  
may appear to lose part of the aura traditionally accompanying it.

To conclude, one must be cautious  in applying  the concepts such as linearity or unitarity of the time  evolution, valid  in the context of pure states, to the measurement processes  where entanglement 
 between the microsystem and the measurement device as well as with the whole world,   
 plays an essential  role.  Their  consequences in these, complicated mixed states may well   (and indeed,  do) look differently from what one is accustomed to from  the 
 experiences  in  pure quantum states kept in isolation.

A few additional remarks on curious subtleties in the concept of unitarity and quantum mechanical predictions are  in Appendix \ref{subtle} for the interested reader.

 \subsection{Degenerate eigenvalues  \label{degenerate} }
 
 We have assumed in  (\ref{stateBisbis}) that the eigenvalues $f_n$ are non degenerate.  Existence of one or  more $f_n$'s  which are degenerate, signifies  that 
 there are some variable(s) compatible with $F$, and the state is characterized, not only by the eigenvalue $f_n$ of $F$ but also by the eigenvalues of those other variables. 
 For simplicity, let us assume that there are two compatible variables,  $F$ and $G$,  $[F, G]=0$,  and that  the state  $\psi$ is given by 
 \be     |\psi\ckt =  \sum_{n,\ell}  c_{n,\ell} \,  | n, \ell \ckt\;, \qquad  F  | n, \ell \ckt = f_n | n, \ell \ckt \;, \quad  G | n, \ell \ckt = g_{\ell} | n, \ell \ckt \;,\label{stateTris} 
 \ee  
 \be   \sum_{n,\ell}     |c_{n,\ell}|^2   =1\,,
 \ee
 instead of    (\ref{stateBisbis}).

 Now   $ | n, \ell \ckt\;$  is the state in which the variables $F$ and $G$ have definite values both;   it is the standard understanding  that  these two variables are simultaneously measurable. 
 Let us consider a simultaneous, double measurement of  $F$ and $G$.   Even though each measurement is a spacetime event,   the contemporaneity of the measurements means that they both refer to the same  micro system, described by the wave function,  (\ref{stateTris}).
 
 The meaning of the contemporaneity of the two compatible measurements,  is illustrated well in the example of the system composed  two spin $\tfrac{1}{2}$ particles,  
in the total spin $0$  state, (\ref{Bohm}), (\ref{etcetc}),  which will be discussed  later  in  Sec.~\ref{entanglement}  in relation to the EPR paradox and entanglement. Let us take 
$F=  s_{1\, z}$ and  $G= s_{2\, x}$  which commute with each other.  In  
  measuring  $F$ and $G$   simultaneously,  the experimentalists must make sure that they are studying the two spin $\tfrac{1}{2}$ particles, 
 from the same, single decay event - the coincidence check. This guarantees  that the two experiments  are done on the same state $\psi$;   the exact contemporaneity of the two 
 measurements (spacetime events)  is immaterial. 
 
 Assuming that such simultaneous measurements are done properly, the result is the wave-function collapse,  
 \be     \sum_{n,\ell}  c_{n,\ell} \,  | n, \ell \ckt\  \otimes |\Phi_0\ckt    \Longrightarrow     |{\tilde \Phi}_{m, k} \ckt \;,      \label{JumpSim}
 \ee
 which generalizes   (\ref{JumpCor}),   where $|{\tilde \Phi}_{m, k} \ckt $   
 is the state of the entangled microsystem-measurement device, labelled by the  double reading,  $F=f_m$, $G=g_{k}$.

  [${\mathbbm Q}$]: ~     What  if only one of  the variables, say  $F$,  is measured in the state,  (\ref{stateTris})?   
 
   In an ideal class of experiments,  the measurement of $F$  might  leave undisturbed the state with respect to $G$.  In the instant $F=f_n$ is measured,   the wave function collapses as,
 \be     \sum_{n,\ell}  c_{n,\ell} \,  | n, \ell \ckt\  \otimes |\Phi_0\ckt    \Longrightarrow    |\psi_G\ckt_n   \otimes   |{\tilde \Phi}_{n}   \ckt \;,      \label{JumpSim}
 \ee
 where
 \be     |\psi_G\ckt_n  =  \sum_{\ell}  c_{n,\ell} \,  | n, \ell \ckt \;, \qquad ({\rm no ~ sum ~ over  } \, \, n )\;. 
 \ee
 A simple example of this kind of experiment \footnote{This class of experiments are known as ``strongly repeatable"   in the literature. }    is the  measurement of  $s_{1\, z}$  in the example  of a system composed of two spin $\tfrac{1}{2}$  particles,   mentioned above. The measurement of  $F=s_{1\, z}$  does not interfere with the state of  $G=  s_{2\, x}$:   the instant  $F$ has been found in the state  $ s_{1\, z} = +\tfrac{1}{2}$,   the system will be  left with a nontrivial wave function   (see (\ref{etcetc})),  
 \be      |\psi\ckt_G=    \frac{1}{\sqrt{2}}   ( |+\ckt -    |-\ckt ) \;,  \qquad       s_{2\, x}   | \pm \ckt  = \pm   \frac{1}{2}   | \pm \ckt \;.
 \ee
 
 On the other hand,  in more general classes of experiments,  the measurement of $F$  may  influence the state with respect to $G$.    In the  simple position determination  at the final stage of Tonomura's  double-slit  experiment \cite{Tonomura},   the measurement of $F= x$, the horizontal  coordinate relevant to the interference fringe  (Fig.\ref{TonomuraExp}),  is  necessarily accompanied by the  determination of $G=y$  (the vertical position of the electron arrival). In other words, the measurement of $F=x$ induces 
 the wave-function collapse both in 
 $F=x$ and  in $G=y$.  
  
  These two  simple examples  are sufficient to show that   $[{\mathbbm Q}]$ is  not a sort of question which can be answered based on some general, abstract principles.  
   The answer depends on the details of the measurement settings, as well as  on the nature of the variables.

\subsection{Macroscopic quantum-mechanical systems \label{MacroQM}}

Sometimes  discussed in relation with the measurement processes are 
the so-called  macroscopic quantum states.  Let us at once note that there are, roughly, 
 two classes of such systems which are to be distinguished \cite{Leggett}.  To the first class  belong many well-known
  phenomena such as superfluidity and superconductivity,  BE condensed ultracold atoms, and actually, many familiar phenomena in quantum field theories, such as spontaneous symmetry breaking, Higgs mechanism, and so on. The mechanism under which a macroscopic (or an infinite) number of identical bosons, elementary or composite, occupy the same quantum states, and form macroscopic wave functions, is well understood.  
 
 A second class of  macroscopic (or mesoscopic) quantum states refer to a large number of particles, or a large molecule,  which behave quantum-mechanically, exhibiting the phenomena
 of diffraction, interference,  and  tunnelling between different macroscopic (mesoscopic) states.  This is a fascinating, rapidly developing experimental research area.

A double-slit experiment by using the molecules  $C_{60}$ shows an example of such a mesoscopic system \cite{C60};  the pair of the SQUID states with macroscopic fluxes of opposite signs, with possible tunnelling between them,  are  an  example of this type of macroscopic quantum systems \cite{Leggett}.  A recent construction of a  mechanical resonator
\cite{Aaron}  kept at $T\sim 20$ mK
is considered to be a promising start for realizing  in the laboratory an analogue of the  linear superposition of the ``dead and alive cat" states.  

None of these phenomena, however, are directly relevant to the question of the measurement problems. The scope of the experimental apparatus is, in a sense, exactly opposite.   The task of an measurement device is to 
pick up the information about the fluctuating  quantum mechanical system, and to convert  its (unique) reading into a {\it  classical state of  matter}.  
The wave-function collapse  (\ref{JumpCor})  is, by definition,  a process in which the coherent superposition of the macroscopic experimental-device states gets eliminated. 

To close a possible loophole in the argument, however, let us note that, 
even though  a generic, classical body can hardly  be characterized as a macroscopic quantum-mechanical system,   
we must still keep in mind that the distinction between the  two  
(i.e., a ``macroscopic quantum state"  and  a classical state)   
 may not always be sharp.  A familiar example is light, which may be regarded as a wave of the classical electromagnetic fields, but it is also a macroscopic quantum state 
describing the collective motion of photons. A magnet can also be seen as a classical state of matter, but its microscopic description leads us to a quantum-mechanical system consisting of macroscopic number of aligned  spins.  It is well-known that in a system of infinite degrees of freedom in two or more dimensions, a tunnelling between two degenerate
vacua is not possible. The system chooses one of them as the unique vacuum.   Spontaneous symmetry breaking is a phenomenon of this kind.

 It could be tempting to try to relate the classical uniqueness of the reading of the measurement device  to spontaneous symmetry breaking \cite{Niew}.   However, typical quantum measurements such as the Stern-Gerlach experiment (and many others) do not seem to be of this nature. 

Our arguments here, on the other hand,   
tell  nothing  about (or against)  the possibility of finding, or fabricating,  macroscopic or mesoscopic quantum mechanical states  such as those discussed  in \cite{Leggett,C60,Aaron},
outside the context of quantum measurement problems.

\subsection{Summary: the main result of this work  \label{Main} } 

Summarizing Sec.~\ref{measurement},   the crux of  the absence of coherent superposition of different terms in the measuring process,  (\ref{step111}),  is the lack of the overlapping  spacetime supports  in the associated wave functions.  As a result of entanglement between the microsystem described by the wave function $\psi$ and the experimental apparatus (and the environment),  the whole system 
gets  rapidly transformed into a mixed state characterized by the diagonal density matrix,   (\ref{diagonal}).   The result of each measurement is the state-vector reduction,
(\ref{JumpCor}).  The experiments on the state $\psi$, (\ref{thestate}),   produce  various results $F=f_n$
with relative frequencies,    $|c_n|^2$.

These discussions indeed integrate  the arguments of  Sec.~\ref{General}  (see Sec.~\ref{SumSec2} for the summary),     where the connection (identification)  between theoretical expectation values and the average 
experimental results were only an assumption.  

Combining the results of  Sec.~\ref{measurement}   (especially,  Sec.~\ref{loss}) and Sec.~\ref{General}, indeed, we have been led to the standard quantum mechanical prediction, 
  that measurements on the state $\psi$ give various results  $F=f_n$ with  {\it relative frequency}  ${\cal P}_n = |c_n|^2$.   
  ${\cal P}_n$  is equal to 
$P_n = |c_n|^2$,  the {\it probability}  that   
 a single measurement gives   $F=f_n$, according to Born's rule.
  
The difference between relative frequencies and probabilities might sound a semantic question,  but it is not.   Most  importantly, what is lacking in Born's rule  is  the  {\it   explanation }     (i)   {\it how} the  knowledge about quantum fluctuations  (the wave function)
gets transcribed, through the measurement processes, into  the   {\it relative frequencies }  ${\cal P}_n = |c_n|^2$ for the different outcomes  $f_n$, and  (ii)   {\it why} and {\it how}  the state-vector reduction 
takes place, after each measurement.  
 
To the best of the author's knowledge, the discussions in this work represent the first significant steps  towards filling  in these gaps, and point towards a
more natural interpretation of quantum mechanical laws. 
 See  Sec.~\ref{Conclusion} for further  discussions.

The rest of the work,  Sec.~\ref{jump} $\sim$ Sec.~\ref{entanglement},  is dedicated to  a few illustrations of how well-known apparent paradoxes and puzzles  in the literature can be naturally  explained.

 \section{Quantum jumps and metastable states \label{jump}   } 

A question   often debated is how to reconcile the  smooth  time evolution of the wave function described by the  Schr\"odinger equation  
with   sudden 
``quantum jumps",  occurring in radioactive nuclei  ($\alpha$ and $\beta$ decays),  the spontaneous emission from the excited atoms, and so on.  

Let us consider an $\alpha$ decay from a metastable nucleus,  $X$, 
\be       {}^A_N \left( X  \right)    \Longrightarrow       {}^{A-4}_{N-2} \left( Y  \right)  + \alpha\;,
\ee
where $N$ is the atomic number and $A$ is the mass number.   By considering some population of the  nuclei $X$,   it seems  natural to write the wave function for the 
system as  
\be        | \psi \ckt  =     | X \ckt    +    | Y \ckt  | \alpha  \ckt\;.   \label{nucleus}  \ee
 We do not bother to write the coefficients in front of  the two terms: 
their norms are not conserved due to the decay, see below, (\ref{smooth}).  
Without going into detail, one can immediately note that the first  ($ | X \ckt$)    corresponds to a bound state, whereas the second  ($ | Y \ckt  | \alpha  \ckt$) describes  an unbounded system:  
 $\alpha$ is a free particle flying away.      Therefore the overlap between the two terms  (the interference) is absent.  In spite of a writing which may indicate otherwise,    (\ref{nucleus}) is not the wave function of a pure state.  It is a mixed state.

In many systems of interest such as this,   there are certain effects  (say, the interaction Hamiltonian, $H_I$), which may be regarded as small, and can be treated as perturbations.   Examples are: 
the electroweak interactions or tunnel effects in nuclei,   the interactions with radiation fields in atoms, and so on.  By first considering the 
system  with the  Hamiltonian $H_0$,    without these interactions,    and taking  $H_I$  into account  appropriately as perturbations, and perhaps with the aid of the analytic continuation for the solution of the Schr\"odinger equation 
with smoothly varying parameters,   
it is possible to define  
the wave function of the metastable state,  $| X \ckt$   (see for instance Chap. 13 of \cite{KKGP}).
The state being metastable,   
   its  energy has a small imaginary part, 
\be   E=  E_R -  i \, \frac{\Gamma}{2} \;, 
\ee
where $\Gamma$ represents the total decay rate per unit time  (or the level width), and  $\tau =   \hbar / \Gamma$ is  the  mean lifetime.   The wave function has a time dependence,
\be   {\tilde \psi}_X(t)  =     e^{- i E t / \hbar}  {\tilde \psi_X(0)  } =       e^{- i E_R t / \hbar}   e^{- \Gamma t /2 \hbar }   {\tilde \psi_X(0) }  \;.       \label{smooth}
\ee

Now how can one reconcile such a smooth time dependence with  sudden  quantum jumps, such as  $\alpha$ decay, a spontaneous emission of photons from an excited atom?  This sort of question, probably mixed up with a philosophical  confrontation between Heisenberg's  (based on the matrix mechanics, with emphasis on the transition elements)   and Schr\"odinger's    (with the wave function obeying smooth differential equations) views   of quantum mechanics,   dominated the earlier debates on quantum mechanics \footnote{In fact, the two questions must be distinguished. The latter, more philosophical ``puzzle" was solved via  the proof of equivalence of the Hilbert spaces  $\ell_2$  and  $L^2$ by Schr\"odinger himself. See e.g.,  Tomonaga \cite{Tomonaga}:   it is not the subject of the discussion here.  }.

A little thought tells us that the  nature  (hence, the solution)  of the problem is basically the same as that of the  measurement problems  considered in Sec.~\ref{measurement}.  The detection of these quantum jumps, e.g., the observation of an $\alpha$ particle suddenly appearing and leaving a track in a Wilson chamber,  is a measurement  as any other.  
Of course,  in general,  a spontaneous decay of a metastable nuclei, atom or molecule, occurs  even without a calibrated measuring device waiting for the decay products to appear.   
But  there is a common feature shared with the measurement processes. The quantum jumps  are  caused by some elementary interactions  in $H_I$   
  such as the photon emission by the electron in an excited atomic level, 
 a Fermi interaction of a neutron, or  an $\alpha$ particle hitting the Coulomb barrier inside a nucleus, which are  all   spacetime  events.  
In this sense,  these pointlike, eventlike  processes capture one of  the fluctuating states  in the original metastable system,   exactly as  in a measurement process  discussed in Sec.~\ref{measurement}.   As such, 
the exact timing of a spontaneous quantum jump,  as well as  the specific details of the final states in each decay,  cannot be predicted. They are manifestation  of the quantum fluctuations of the system.   Their averages are, however,  
 described by ${\tilde  \psi}$.
 
 There is, however,  a crucial
difference   between  the general measurement processes discussed in Sec.~\ref{measurement} and the spontaneous, quantum jumps.  It  is  the fact  that  in the former  case,   the wave function $\psi$ in (\ref{stateBisbis})  describes  a  
true  quantum state.  {\it It never spontaneously collapses  (i.e., decays)   to one of the  fluctuating states,  if left undisturbed}.
In  a measurement process, it  is the interactions with  macroscopic, measuring instruments, sensitive to the microscopic state described by $\psi$, that pick  up one of the possible fluctuating states,  experiment by experiment,  making the microsystem  appear to  jump  into one of the eigenstates, as in     (\ref{JumpCor}).

From the very way the wave function $\tilde \psi$  for a metastable state is defined,  where  $\Gamma$ represents the total decay rates of all possible decay processes of the parent particle,  
it is quite clear  that   $\tilde \psi$   represents  an effective description of the metastable ``quantum state", in which the coarse-grained time dependence (decay events)  has been smoothed out. It does not have the same status as the wave function of a genuine  quantum state $\psi$, (\ref{stateBisbis}).  

In conclusion,  the conundrum of apparent  impossibility of  reconciling  the smooth time-dependent Schr\"odinger equation with quantum jumps,   appears to have been caused by  the confusion between  the concept of the true wave function  $\psi$  and
that of  an effective ``wave function" for a  metastable ``state", ${\tilde \psi}$.

\subsection{Atomic transitions: other examples  of quantum jumps }

Spontaneous emission of photons from an excited atom is a typical quantum jump, as discussed above.  Just to be complete, let us note that the two other kinds of atomic 
transitions,  absorption-excitation  and stimulated emission,  are  also both eventlike processes, sharing common features with the decay of  metastable states, as well as  
with the measurement processes. Note, for instance,  that in the photoelectric effect,  absorption of the photon and excitation-ionization of the  $Li$  atoms in the target plate,  
is an eventlike process, serving as a measurement of the incoming-photon energy \footnote{In passing,  this example shows that a statement sometimes found in the literature,  ``every measurement reduces to a position measurement",   is another unfounded myth in the quantum-measurement discussions.}.

\section{Schr\"odinger's cat  \label{cat}}

In the discussion involving Schr\"odinger's cat,   the initial ``wave function"    is  a combination  (\ref{nucleus}),  between the undecayed nucleus  $|X\ckt$
and the state after decay,   $ | Y \ckt  | \alpha  \ckt $.     To maximally simplify the discussion and to go directly  into the heart of the problem, let us eliminate altogether  the intermediate,  diabolic device which upon receipt of the $\alpha$ particle leads to the poisoning of the cat, and treat the cat  directly   as the measuring device, $|\Phi \ckt$.  When it detects the $\alpha$ particle, it dies;  when it does not, it remains alive. There are    
two terms in  the measurement process,   (\ref{step000}), (\ref{step111}),  which here reads 
 \bea   && | \psi \ckt \otimes |\Phi_0\ckt   =   \left(   | X \ckt    +    | Y \ckt  | \alpha  \ckt  \right)    \otimes |\Phi_0\ckt  \label{step00}  \\
 &  \longrightarrow   &      | X \ckt     \otimes |\Phi^{(alive)} \ckt     +      | Y \ckt  | \alpha  \ckt     \otimes    |\Phi^{(dead)} \ckt        \label{step11} \;
 \eea
 (we dropped also the rest of the world, $| Env \ckt $). 
Such a process appears  to lead to the superposition of the  dead and alive cat, which is certainly an unusual 
notion, difficult to conceive, to say the least. 

Actually, there are at least three elements  (of  abuse/misuse  of concepts)    in this argument, each of which individually invalidates it,  eliminating the notorious conundrum.

  The first is the fact, as seen in the last section, that  the ``state"  of the undecayed nucleus, $X$,  is not a pure state.   The  ``wave function"   ${\tilde \psi}_X$ describing it is not a proper wave function but an effective one, in which the coarse-grained time dependence has been averaged out.    Second,  the linear superposition 
$ |\psi \ckt  \sim   | X \ckt    +    | Y \ckt  | \alpha  \ckt $,   as  discussed around (\ref{nucleus}),    does not represent a pure
state (even if we ignore the problem of $ | X \ckt $  itself),  but a mixed state:   $ | Y \ckt  | \alpha  \ckt $  represents the decay product of  
 $ | X \ckt $, unbounded and incapable of interfering with   the latter.

 Finally,  even putting aside these two  issues,   the process  (\ref{step00}),   (\ref{step11})   has clearly all the characteristics of a general  measurement process discussed in Sec.~\ref{measurement}.
As explained  there,    the 
 wave functions describing the two terms of (\ref{step11}) lack  overlapping spacetime supports.  
  There are no interferences between the terms involving the coupled system  involving  the microsystem  and the classical  measurement device (the cat here). The formula  (\ref{step11}) represents  a mixture.  
 
Equivalently,   and  perhaps more intuitively,    one can  use the notion of the wave-function collapse. 
The exact timing of the spontaneous  $\alpha$ decay cannot be predicted, as it reflects a quantum fluctuation. But the moment  an  $\alpha$ particle is emitted,  and the cat gets hit by it  (the instant of the measurement), the wave function collapses,  
 \be   | \psi \ckt \otimes |\Phi_0\ckt     \Longrightarrow      | Y \ckt  | \alpha  \ckt     \otimes    |\Phi^{(dead)} \ckt  \;.  
 \ee

 Recapitulating,  Schr\"odinger's cat paradox,  as it has been formulated originally,  
was concocted by improper use of concepts such as the superposition principle, unitary and linearity of evolution, and  mixing up the pure and mixed states.  Any apparently paradoxical feature disappears, once these errors are taken into account   \footnote{ What happens actually is simple: as long as the nucleus has not decayed (no $\alpha$ particle emission) the cat remains alive. The exact timing of the $\alpha$ decay cannot be predicted. But the moment the $\alpha$ particle is emitted, and the cat gets hit,  it dies. 
That is  all.  
There are no problems in describing this process appropriately  in quantum mechanics, as seen above. }.

This being so, it is an entirely different question whether or not   
 quantum-mechanical superposition of macroscopic states (a distant analogue of the superposed dead and alive cat),  as those discussed  briefly in Sec.~\ref{MacroQM},
 can be  somehow 
realized experimentally.

   \section{Quantum entanglement  and  the EPR ``paradox" \label{entanglement}}

Let us consider  now  the famous Einstein-Podolsky-Rosen setting (in Bohm's version) of a total spin $0$ system decaying into two spin $\tfrac{1}{2}$ particles
 flying away from each other.  
The (spin) wave function is given by 
\be     |\Psi_0\ckt =  \frac{1}{\sqrt 2}   \left(   |\!\uparrow\ckt |\!\downarrow \ckt  -   |\downarrow \ckt   |\!\uparrow\ckt       \right)      \label{Bohm}
\ee\!
where   $  |\!\uparrow\ckt$  ($ |\!\downarrow \ckt  $) represents the state of the 1st and 2nd  spins,  $s_{1, z}= \pm \tfrac{1}{2}$,  $s_{2, z}=\pm \tfrac{1}{2}$.    
As spin components of the single spins  do  not commute with the total spin  $  {\mathbbm S}_{tot}^2$,  e.g., 
\be   [   {\mathbbm S}_{tot}^2,   s_{1, i}]    \ne 0\;,\quad   [   {\mathbbm S}_{tot}^2,   s_{2, j}]    \ne 0\;,\qquad    i,j=x,y,z\;,
\ee
$s_{1\, z}, s_{2\, y} $, etc.,  are  fluctuating between $\pm \tfrac{1}{2}$ in the state $\Psi_0$.   Note that  $\Psi_0$ can be written in infinitely many different  ways, reflecting possible fluctuation modes of the subsystems, 
\be     |\Psi_0\ckt =   \frac{1}{\sqrt 2}   \left(   |+\ckt |- \ckt  -   |- \ckt   |+\ckt       \right)    =   \frac{1}{2}   \left[  \, |\!\uparrow\ckt (  |+\ckt -    |-\ckt )  -   (  |+\ckt -    |-\ckt )    |\!\uparrow\ckt   \,   \right]    = \ldots \;.\label{etcetc}  
\ee

It is of fundamental importance to realize that  such (correlated)  fluctuations of the two subsystems     -   quantum entanglement -    are present  however distant the subsystems may be, and even if they might be relatively spacelikely  separated. 
  Nothing in $\Psi_0$  tells  us otherwise.  Nothing in quantum mechanical law  says that  such entanglements  are 
present only when the subsystems are nearby, e.g.,  less than  $10^{-8}$ cm,  or   less than $10^{+28}$ cm.  Simply,  {\it  quantum mechanics does not contain any fundamental parameter with the dimension of a length} \footnote{String theory, or quantum gravity, does have one, of the order of  the Planck length,  $\ell_P \sim10^{-33}$ cm.
We will not discuss  here either possible genuine modifications of quantum mechanics at such regimes, the issues of information-loss paradox and blackhole entropy, or  the consistent formulation of quantum gravity. Still, we do not adopt the view that either of them  (or   $\ell_P$)    might have any relevance to the measurement problems  being  discussed in this work.}.

In the EPR-Bohm experiment, this means that an experimental result at one arm,  say, $s_{1\,z}=  \tfrac{1}{2}$, might appear to   imply    ``instantaneously"  that the second experiment at the other arm would be in the state   $s_{2\,z}= - \tfrac{1}{2}$, even before 
it is actually performed, and however distant they may be.   This could sound paradoxical (sometimes the term ``quantum nonlocality"  is used).  The second experimentalist, not having access to the first experiment, might  have (should have?)  expected to find  the results  $s_{2\,z}= \pm \tfrac{1}{2}$,  a priori  with equal probabilities.   

 This kind of argument  has  led to the introduction of  the hidden-variable hypothesis  (see \cite{Bell} for discussions), although such a deduction was not actually justified.  
We limit ourselves here to noting that this argumentation  had several logical  flaws.  First, the contemporaneity of the two experiments is not really an issue: the experimentalists must just ensure that they are studying the two spin $\tfrac{1}{2}$ particles  from the same decay event -  the coincidence check. The exact time ordering, which depends on the 
reference system chosen,  cannot matter.      In any case, as the two subsystems are spacelikely separated, it is untrue that the second experimentalist would  know   instantaneously   the result of the first experiment,  or actually, even whether or not  the first measurement  has been indeed performed.     The information transmission is itself a dynamical process. 
Finally,  the two ``simultaneous" experiments would capture only those states of the two subsystems  
fluctuating according to the entangled wave function, (\ref{Bohm}),   (\ref{etcetc}). In conclusion,  for the second experimentalist the system presents itself as  a mixture with an  unknown density matrix.  {\it    He (she) simply does not know what to expect. }  There was no paradox.

Coming back to physics,   the first experimental result  $s_{1\,z}=  \tfrac{1}{2}$  {\it  does imply,  instantaneously}   (whatever it may mathematically  mean),  that the other spin is in  the state   $|{\downarrow}\ckt
= \tfrac{1}{\sqrt{2}} ( |+\ckt -    |-\ckt )$. This follows from  the explicit form of the wave function  (\ref{Bohm}),  (\ref{etcetc}) \footnote{This  quantum nonlocality, an aspect of quantum entanglement, is real.   But it should not  be confused with the dynamical concept of locality (or causality).   Quantum mechanics is perfectly consistent with causality and locality,  due to the fact that all the fundamental, elementary interactions are local interactions in spacetime  (see Appendix.~\ref{Minkowski} and \cite{Peskin}).   No dynamical effects, including the information transmission,  propagate faster than the velocity  of light.  
  It is perhaps best to regard (\ref{Bohm})  simply as a particular kind of macroscopic  quantum-mechanical state   (cfr., Sec.~\ref{MacroQM}).}.
  But unless  the second measurement is done,     any components of the second spin  other than $s_{2\, z}$  are  still  fluctuating \footnote{This  point  shows that  the often mentioned 
  ``Bertlmann's socks" classical action-at-distance analogy,  is not valid.  Quantum non-locality is subtler,  and is  different  \cite{Bell}.  }.   Thus   if the
two Stern-Gerlach type experiments are performed at the two ends simultaneously, by using the magnets directed to generic directions ${\bf a}$ and ${\bf b}$,   
they would find,  experiment by experiment,   apparently random results, such as  $(s_{1\, a}, s_{2\,  b})  =    (\tfrac{1}{2}, \tfrac{1}{2}),  (\tfrac{1}{2}, - \tfrac{1}{2}),    (-\tfrac{1}{2}, -\tfrac{1}{2})$, etc.
But their fluctuation average is encoded in $\Psi_0$:
\be     \brc \Psi_0 | s_{a}  s_{b} | \Psi_0  \ckt =  -   \frac{1}{4}  \cos \theta_{a,b} =    -   \frac{1}{4} \,   {\bf a}\cdot {\bf b}\;.\label{QM}
\ee

Renowned inequalities  by Bell (1960) \cite{Bell}   for this system,  and   CHSH  inequalities \cite{CHSH}  
for analogous,  polarization correlated photon pair  experiments,  show that hidden-variable models, independently of the details,  cannot 
simulate completely quantum mechanical predictions based on formulas such as (\ref{QM}),  which encode  the entanglement information.   Beautiful experiments by Aspect et. al.  (1981) \cite{Aspect} have subsequently demonstrated that,  whenever  the hidden-variable alternatives  and quantum mechanics
give discrepant predictions (for certain sets of magnet orientations or polarizer axes), the experimental data confirm
quantum mechanics, disproving the former.

See also  Chiao et. al. (1995) \cite{Chiao}  for a series of related experiments and discussions.

\subsection{Entanglement  among more than two particles  \label{Mermin}}

An amusing variation of the EPR-Bohm experiment to the one with more than two particles was considered by Mermin \cite{Mermin}, which    (apparently)    shows a further   mysterious and paradoxical feature of quantum entanglement \footnote{The discussion of this section has briefly appeared in  Sec.~18.3 of  \cite{KKGP}.}.
Consider
   three spin-$\tfrac{1}{2} $  particles in the  state
\be \ket {\Psi_{M}} =  \psi({\bf r}_{1}, {\bf r}_{2},{\bf r}_{3}) \,  \frac{1}{\sqrt{2}}  \left(  \ket {\uparrow}\ket {\uparrow}\ket {\uparrow} - \ket {\downarrow}\ket {\downarrow}\ket {\downarrow} \right)\,,
\label{Mermin0} \ee
 with the three particles receding from each other (e.g., they are the decay product of some parent particle), each pair getting separated by a spacelike distance.   Let us consider four operators
 ($\sigma$'s being  twice the spin operators, $s_{1,x}= \tfrac{1}{2}  \sigma_{1,x}$, etc.),
\be   M_{1} =  \sigma_{1\, x} \,  \sigma_{2\, y}  \sigma_{3\, y}, \quad 
M_{2} =  \sigma_{1\, y} \,  \sigma_{2\, x} \, \sigma_{3\, y}, \quad 
M_{3} =  \sigma_{1\, y} \,  \sigma_{2\, y} \, \sigma_{3\, x}, \quad  M_{4} =  \sigma_{1\, x} \,  \sigma_{2\, x} \, \sigma_{3\, x}\;,  
\ee
which  commute with each other.  
A remarkable property of state (\ref{Mermin0})  is that it is a simultaneous 
eigenstate of all four operators,
\be M_{1}\,  \ket {\Psi_{M}} = M_{2}\,  \ket {\Psi_{M}} =M_{3}\,  \ket {\Psi_{M}} =\ket {\Psi_{M}} \,, 
\label{Mrelation1}   \ee 
and 
\be  M_{4}\, \ket {\Psi_{M}} = - \ket {\Psi_{M}}\;,
\label{Mrelation2}  \ee
as can be easily verified. 
Now consider measuring $M_1$:  simultaneous measurements of commuting variables $\sigma_{1\, x}$,   $\sigma_{2\, y}$  and $\sigma_{3\, y}$.  Each of them will yield $\pm 1$ as a result: let us call  them $m_{1\, x}$, $m_{2\, y}$,  $m_{3\, y}$, respectively. Because of the first of  (\ref{Mrelation1}), the triplet of the outcomes will satisfy
\be   m_{1\, x} \, m_{2\, y} \, m_{3\, y}  =  1\;.  \label{Mermin1}
\ee
Similarly, other triple experiments  for  $M_2$, $M_3$, and $M_4$ 
will give  results,
\begin{subequations}\label{totmermin}
\begin{align}
m_{1\, y} \, m_{2\, x} \, m_{3\, y}  &=  1\;,  \label{Mermin2}\\
 m_{1\, y} \, m_{2\, y} \, m_{3\, x}  &=   1\;,  \label{Mermin3}\\
 m_{1\, x} \, m_{2\, x} \, m_{3\, x}  &=  - 1\;.  \label{Mermin4}
\end{align}
\end{subequations}
Now, as each of the $m$'s take only values $\pm1$,  and because of the crucial minus sign in 
(\ref{Mermin4}),  the results  (\ref{Mermin1}), (\ref{Mermin2}), (\ref{Mermin3}), (\ref{Mermin4})  are mutually incompatible!  Is quantum mechanics inconsistent? 

Actually, we have been a little too hasty in jumping to a wrong conclusion. We incorrectly regarded  $m_{3\, y}$ appearing in (\ref{Mermin1}) and  in (\ref{Mermin2}) as the same number, but this is not at all guaranteed. Similarly for  $m_{1\,x}$  in  (\ref{Mermin1}) and in  (\ref{Mermin4}).    The point is that  the triple measurement for $M_1$  (which yields (\ref{Mermin1}))  and another  triple experiment for $M_2$   (which gives $m$'s satisfying (\ref{Mermin2})), for instance,  cannot be performed simultaneously,  {\it in spite of} the fact that the operators  $M_{1}$ and $M_{2}$ commute.   As  $[\sigma_{1\, x}, \sigma_{1\, y}] \ne 0$,  the experimentalist  at site 1 cannot measure them simultaneously.
 More generally,   even though the four operators $M_{1}, \ldots,  M_{4}$  commute with each other,   observation of such 
composite variables, each of which requires three ``simultaneous'' measurements taking place  in different sites,  cannot be simultaneously performed,  against  the standard lore of quantum mechanics \footnote{That is,  ``commuting operators correspond to compatible observables, which can be measured simultaneously."  This assertion  is simply wrong here. }. 
Stated more precisely,  relations  (\ref{Mermin1}), (\ref{Mermin2}), (\ref{Mermin3}), (\ref{Mermin4}) necessarily refer to four different (triple-measurement-) events, with unrelated $m$'s.
 They should have been written as 
 \be  m_{1\, x} \, m_{2\, y} \, m_{3\, y}  =  1\;,   \quad 
m_{1\, y}^{\prime}  \, m_{2\, x}^{\prime}   \, m_{3\, y}^{\prime}    = 1\;,  \quad 
 m_{1\, y}^{\prime \prime}    \, m_{2\, y}^{\prime \prime}  \, m_{3\, x}^{\prime \prime}   = 1\;, \quad 
 m_{1\, x}^{\prime \prime \prime}  \, m_{2\, x}^{\prime \prime \prime}   \, m_{3\, x}^{\prime \prime \prime}  =  - 1\;.  \label{Mermin44}
\ee
There is  no inconsistency whatsoever,   among   the four results.
       
   This  system was presented in \cite{Mermin} as a case in which the discrepancy  between a  hidden-variable  theory and quantum mechanics can be seen more sharply than in  the EPR
   experiment/Bell's inequalities  where  statistical results of many measurements are needed to discriminate between the two types of theories.   The idea would be that 
   the  particular set of the hidden variables  $\lambda$, supplementing $\Psi_M$, which give rise to the results     (\ref{Mermin1}), (\ref{Mermin2}), (\ref{Mermin3}), would necessarily
   predict $ M_4 =  m_{1\, x} \, m_{2\, x} \, m_{3\, x} =  +1$,  contradicting  the quantum mechanical prediction,   (\ref{Mermin4}),   with a single experiment!
      
  Actually, such an argument neglects the quantum mechanical aspect of this process we discussed above.   
   The four triple measurements $M_1$,  $M_2$, $M_3$, $M_4$     necessarily refer to four distinct experiments  (four different three-particle decay events), and the state $|\Psi_M\ckt $
   is such that the results of these experiments satisfy  (\ref{Mermin44}), each of  $m$'s, $m^{\prime}$'s,  $m^{\prime \prime}$'s and  $m^{\prime \prime \prime}$'s being  equal to $\pm1$.
  Any plausible hidden-variable model must be such that at least these conditions are correctly simulated.  
   It would not be surprising to find a result which contradicts quantum mechanics, if an assumption violating the principles of quantum mechanics were used to get  it  
\footnote{ Bell's criticism \cite{Bell}  on von Neumann's ``proof" of the impossibility of  the dispersion-free states  (theory of hidden variables), is of similar nature. 
 An analogous comment can be made on the (improper)  use of the  Kochen-Specker theorem, as a way to disprove the hidden-variable theories.   Strictly speaking, a similar criticism applies even  to Bell's inequalities, or on Bohm's pilot-wave theory (a particular hidden-variable
 model), as well. They all contain some assumptions which violate quantum mechanical laws.   It  is no mystery why these 
 alternatives fail to reproduce quantum mechanics fully.    See Sec.~19  and  Sec.~23.1 of  \cite{KKGP}.     }.

\section{Universe and quantum measurements \label{Universe}}

Before concluding,  let us briefly wander into a few general but related issues which our considerations on quantum measurements have illuminated.   
Reviews on the concepts of pure and mixed states, and related questions of factorizatioin versus entanglement
 in Sec.~\ref{mixed} and Sec.~\ref{factorization}  have brought us to think of the universe as a large (today) single quantum system (``the wave function of the universe"), which is  in an entangled, mixed state.
   
Massive bodies  in it  (galaxies, stars, planets, objects on Earth) generally behave classically  (i.e., as mixed entangled systems  characterized by decoherence  \cite{Zurek,Joos}.)
Pure quantum states however occur in Nature (i.e., in our universe)  in the form of cosmic rays in the interstellar (intergalactic)  spaces: they are  free particles such as  neutrinos, photons and ultra high-energy protons \footnote{There are also exceptional, macroscopic, quantum mechanical systems such as neutron stars. }.
 They are produced in astrophysical phenomena, stars, supernova, etc.,  and travel freely, except under adiabatic effects of gravitational and magnetic fields of galaxies.  Cosmic Microwave Background (CMB)  (and dark matter) at $T \simeq 2.7$K  ($T_{DM} =?$) do not disturb them significantly. 

Pure states occur  naturally  also  on Earth,  due to radioactivity  ($\alpha$,  $\beta$ and  $\gamma$ decays)  of unstable nuclei.  Unless  intercepted, these $\alpha$ particles,  electrons and photons,  
are free particles. Their studies indeed led to the discovery of quantum mechanics  between the end of the 19th century and the early 20th century \footnote{Black-body radiation  which played the pivotal role in the discovery of quantum physics by Planck (1900), and CMB discovered later,  are  thermodynamical, statistical mixtures, at a given temperature $T$:  
 they describe free photons whose average energies are specified by $T$. 
 }.  Cosmic-ray physics also played an important role
in the development, for the same reason: they were sources of pure quantum states.  The pure states are today  also artificially  produced and studied  in
isolated pockets  in the world - physics laboratories  on Earth with good vacuum technology and accelerators. 

As discussed in Sec.~\ref{measurement},    quantum measurement is a special procedure   of coupling (entangling) various pure states (particles) produced artificially
in the laboratories,  or   (primary or secondary)  cosmic rays hitting Earth,   to particular kinds of mixed-state bodies (the experimental devices) in entanglement with the environment. The latter has,  typically,  the characteristics of an ionization chamber,  to be triggered by the initial spacetime eventlike interactions  (which capture the fluctuating quantum state of the microsystem under study),   chain-multiply and amplify enormously the result of the initial triggering-interactions.  The measurement result is now in the form of  particular, entangled mixed (i.e., classical) states with the reading of the  measurement result.

The quantum measurement processes may thus be seen as  a very special, controlled  way of reproducing  what happens  constantly in Nature:    the loss of factorization and pure-state nature   of  a subsystem, via interactions (entanglements)  with surrounding background, into a mixture.

\section{Summary and conclusion \label{Conclusion}}

The key observation of this work that the basic entities  (the degrees of freedom) of our world are various types of {\it  particles}, leads to the idea    that each measurement, at its core, is a spacetime, pointlike event.  The crux of the absence of coherent superposition of different terms in the measuring process (\ref{step111})  is, indeed,  the  consequent  {\it  lack of the overlapping  spacetime supports} in the associated wave functions, instant after measurement which is the process of  entanglement between the microsystem and the experimental device and the environment.
 It is in this sense the expression  (\ref{step111}) represents a mixed state.
   But it also implies that the aftermath of each measurement  corresponds to a single term of (\ref{step111}).    
This (also empirical) fact, is  known as the ``wave-function collapse",   or the state-vector reduction.

The discussions of Sec.~\ref{measurement}  have led to the diagonal density matrix, (\ref{diagonal}).  In particular,  
the information encoded  in the wave function    $|\psi\ckt    =    \sum_n  c_n   | n \ckt$  has been  transferred via entanglement with the experimental device and environment into the
relative frequencies for finding the experimental results $F=f_n$,      ${\cal P}_n = |c_n|^2$.  
Combined with the equality between   ${\cal P}_n = |c_n|^2$ (the relative frequencies) and $P_n = |c_n|^2$ (the probabilities), reviewed in Sec.~\ref{General}, our results represent the first steps towards filling the logical gaps in Born's rule, explaining the state-vector reduction, and pointing  towards a more natural interpretation of quantum mechanics.

   Our discussions  do   confirm the standard  Born rule as  a concise way of summarizing the predictions of quantum mechanics.  
But the fact that it  is formulated in terms of the  ``{\it  probability} certain result is found in an experiment",  which might  (or might not) be eventually  performed  - perhaps by an experimentalist with a Ph D \footnote{Borrowed from Bell's remarks, e.g.,   in ``Quantum mechanics for cosmologists" (reprinted in \cite{Bell}).}  -   has  always caused  confusion and conceptual difficulties
 \footnote{Every teacher of a quantum mechanics course knows about a  curious feeling of guilt  he (or she) experiences, on the day of proclamation of the
``fundamental postulate"  (Born's rule).  }.

 In  ultimate analysis, the problem of  Born's rule   is  the fact that it 
presumes a human intervention.

We instead take the quantum fluctuations described by the wave function
  as real,  and propose  that they represent the fundamental laws of quantum mechanics.   They are there, independently of any experiments, and in fact,  whether or not 
  we human beings are around  \footnote{From this new perspective,  the notion of the ``wave function of the universe" makes perfect sense.  See also Sec.~\ref{Universe}.}.   In case  a measurement is done,    the good experimental device  will  faithfully reflect them, giving in general an apparently random outcome each time.  This is as it should be, even though the average-over-experiments-result can be predicted, as the fluctuation averages are encoded in the wave function.

What we propose here  is  a  slightly unconventional way of {\it  interpreting} quantum mechanical laws.  Indeed, the solid textbook materials and the standard methods of analysis  
in quantum mechanics,  all remain unchallenged. 
After all, we know that quantum mechanics is a correct theory.
The new perspective, however,  suggests us to put less emphasis in the discussion of quantum mechanics, on
certain familiar concepts  such as  
 probability,  statistical
ensembles, Copenhagen interpretation,   etc., 
 which historically guided  us, but also caused  misunderstandings and  introduced clouds
in the discussion on quantum mechanics, for many years.

  Accepting the quantum fluctuations described by the wave function as real, 
we simply give up  pretending, or demanding,  that in a fundamental, complete theory describing Nature   the result of each single experiment should necessarily be predictable.

As the known structures of our universe (including  galaxies, stars,  planets and ourselves) are likely to have grown out of uncontrollable quantum density fluctuations at some stage of the inflationary universe, 
as  cosmologists tell us  today,  this is perhaps a healthy attitude to take, to attain the proper  understanding of  the quantum laws of Nature, which have far-reaching consequences in all present and future science domains in physics and beyond, from  biology and chemistry,   quantum information and quantum computing,    and 
possibly, to the brain science.

\newpage

\section*{Acknowledgements} 

The author is grateful to Pietro Menotti  for  many in-depth discussions and for sharing generously his knowledge  and insights   with him,  
without which this work could not have been completed,   and for exchanges of views on 
 the fundamental aspects of quantum mechanics which have been going on  for thirty years.  The author's gratitudes also go to 
Giampiero Paffuti for the invaluable collaboration work  in writing the book \cite{KKGP},  to 
Gabriele Veneziano for explaining some aspects of cosmology,  to Hans-Thomas Elze for discussions, suggestions  and for carefully reading the manuscript, to Matteo Brini for discussions on \cite{Renner}, 
to  the local experimental colleagues  Francesco Forti, Giovanni Signorelli and Alessandro Tredicucci  for conversations on the measurement devices known today, and to Sergio Spagnolo and 
Ferruccio Colombini from the Mathematical Department for help  in mathematics.
 Private communications with (late) Akira Tonomura about his original works   
 had been  instrumental for the formation of the author's ideas about quantum mechanics.  
This work is supported by  the INFN special research initiative grant, ``GAST"  (Gauge and String Theories). 
 
{}

\appendix

\section{Proof of  ${\cal P}_n =  P_n$  \label{Vandermonde}}

Requiring the expectation values of various powers of $F$,    (\ref{fund1}), 
to be equal to the experimental  expected frequency averages,  (\ref{expect}),   for $N=1,\ldots, D$,  one finds
\be         \sum_{n=1}^{D} f_n^N \, ({\cal P}_n -   |c_n|^2) =0\;,  \qquad N=1,\ldots, D\;.
\ee
This can be written as 
\be       \sum_{n=1}^D    V_{mn}  \,  d_n   =0\;, \qquad  d_n \equiv      {\cal P}_n -   |c_n|^2\;, \qquad m=1,\ldots, D
\ee
where  $V_{mn}  \equiv    (f_n)^m $ is a Vandermonde matrix.  As  its determinant 
\be   \det V = ( \Pi_{n=1}^D  f_n ) \, \,      \Pi_{1\le n < m \le  D}    (f_m -  f_n)   
\ee
is nonvanishing, we find 
\be    d_n=  0\;, \qquad    .^.. \quad  {\cal P}_n  =   |c_n|^2 = P_n\;, \qquad  n=1,2, \ldots, D\;.  
\ee

\section{Unitarity and measurements \label{subtle} }

An attentive reader may have noted certain subtleties in the concept of  unitarity and measurements.  
Namely,  different expressions of unitarity,      (\ref{obvious}) for the sum of the relative frequencies,  (\ref{trivial2}) for the normalized sum of all possible experimental outcomes, 
and (\ref{SumP}), the total probability,   all look trivial, each in its own way!   Even the  conservation of the norm of the state may sound suspicious, since a quantum state is defined by the ray of the vector in Hilbert space:   its norm is unphysical. 

What unitarity tells  us is that, once the norm of the state   $\psi$  is fixed by  convention, and the states in the base   $\{| n \ckt\} $  are chosen to be orthonormal,  then the evolution  $|\psi \ckt   \to U  |\psi \ckt  $   and     $ c_n | n \ckt  \to  c_n U   | n \ckt $,      maintain  the norm of each,  $U= e^{- i H t / \hbar}$  being a unitary operator.  It follows that   each $|c_n|^2$,   their sum, $\sum_m  |c_m|^2$, thus the ratio  
$ |c_n|^2 /  \sum_m  |c_m|^2$,   
are all   $t$ independent.   This is welcome, since  verification of the predictions on various probabilities (Born's rule) requires repeated experiments performed at different times. The predictions in terms of the relative frequencies, and the experimental results, refer both directly to the repeated  events occurring at different instants. An identical state $|\psi\ckt$ must be prepared ad hoc each time  (e.g. by using a repeatable experiment), but once prepared,   $|\psi\ckt$  may be allowed to wait, before being sent to the actual experiment region, for an arbitrary amount of time.   The start time should not be essential. This is consistent,  as an experimental test in quantum mechanics involves (in principle) an infinite number of repeated measurements.     For instance, one may throw away the first hundred events  from the collection of Tonomura's  electron images Fig. \ref{TonomuraExp}:   the final result remains the same.

As we saw in Sec.~\ref{measurement},  during the measurements,  where the microsystem gets entangled with the apparatus-environment state, leaving a mixed state
without coherent superposition of the different terms,  unitarity  $\sum {\cal P}_n=1$   remains valid,  
 even though  each experimental result $F=f_n$ refers explicitly to experimental event which  occurred at different time, 
 and each is  accompanied by state-vector reduction.

\section{Quantum mechanics and Minkowski spacetime \label{Minkowski}}
 
Our key observation that each measurement is at its core a spacetime event occurring at some $(x_1, x_2, x_3, t)$,
  explains  the absence of the overlap in spacetime supports in the wave functions corresponding to different terms in (\ref{step111}).  
 This, of course,  does not imply that these variables, including the time $t$,  are treated as  dynamical variables.  In spite of a few existing debates, the time $t$ is not a dynamical variable in quantum mechanics.  

    The spacetime coordinates $(x_1, x_2, x_3, t)$ represent  the frame  of the  Minkowski spacetime  in which all particles move. In particular, the time $t$ is a common (time) coordinate for describing the time evolution of all particles present in the system. Actually,  the same can be said of the space coordinates  $(x_1, x_2, x_3)$,    even if the confusion often made   between them and the position operators (and their eigenvalues) of the particles obscures this fact.   For instance, 
    the wave function of a system of two particles  is a function $\psi({\bf q}_1, {\bf q}_2, t)$ of  six position operators, whereas  the fixed space coordinates are always just 
 $(x_1, x_2, x_3)$.  The latter is the spectrum for both $({\bf q}_1)_{x,y,z} $ and   $({\bf q}_2)_{x,y,z} $.
 The position variables  ${\bf q}_i$'s fluctuate;  $(x_1, x_2, x_3)$ do not.  
 
That there is a basic symmetry  between the space and time coordinates even in nonrelativistic quantum mechanics,  is actually  well known from the quantization rules 
 \be    E    \to    i    \hbar  \frac{\partial}{\partial t}\;, \qquad   {\bf p}  \to   - i  \hbar  \nabla\;, \label{simple} 
\ee
for energy and momentum,  which can be cast in a relativistically covariant form, 
\be      P_{\mu}  \to    i    \hbar  \frac{\partial}{\partial  x^{\mu}} \;, \qquad    x^{\mu} =  (t,  x_1, x_2, x_3) \;.  
\label{struttura}
\ee
Such a structure  is what makes the generalization of quantum mechanics to relativistic quantum mechanics   (the Dirac and Klein-Gordon equations replacing the Schr\"odinger equation)  
and to relativistic quantum field theories  (QFT),
at all  possible.
It is within this last framework (QFT)  that   the standard model of the fundamental (strong and electroweak)  interactions has been successfully constructed.

Without going into detail, let us recall also that three important properties of  quantum mechanics,  the spin-statistics connection,  locality and causality, all require relativistic-QFT considerations  for their proof.   See,  for instance,  Landau-Lifshitz, Vol. 4 \cite{LL4} 
for the first, and Peskin et. al. \cite{Peskin},  for the second and third.

 As a final remark,  we recall that the  simple form (\ref{simple})  of the momentum operator as a differential operator with respect to the canonically conjugate 
position variable, is valid with any coordinate choice, but only among the Cartesian coordinates  but not in  curvilinear ones (see, e.g., Sec.~7.8 of \cite{KKGP} for a review), 
suggesting a key role  the flat Minkowski spacetime plays in the  standard  quantum mechanics framework. 

Important physics arises from quantum mechanics (quantum field theories) in the presence of the spacetime curvature,
gravity and general relativity.  Famous examples are those in  black-hole-physics context (e.g., Bekenstein entropy, Hawking radiation), and also  the problem of structure formation in our universe
 which is becoming an essential element of cosmology today.  
These issues are however beyond (and in our understanding,  outside) the scope of the present work.

\end{document}